\documentclass[12pt]{article}
\usepackage{amssymb}
\usepackage{amsmath}
\usepackage{amsthm}
\usepackage{color}
\usepackage{comment}
\usepackage{geometry}		% allows easy margin settings
\usepackage{graphicx}
\usepackage{hyperref}

\let\originalleft\left
\let\originalright\right
\renewcommand{\left}{\mathopen{}\mathclose\bgroup\originalleft}
\renewcommand{\right}{\aftergroup\egroup\originalright}

\usepackage{caption}
\usepackage{subcaption}

\begin{document}

\newcommand{\cO}{\mathcal{O}}
\newcommand{\cS}{\mathcal{S}}
\newcommand{\rD}{{\rm D}}

\newcommand{\removableFootnote}[1]{\footnote{#1}}

\newtheorem{theorem}{Theorem}[section]
\newtheorem{corollary}[theorem]{Corollary}
\newtheorem{lemma}[theorem]{Lemma}
\newtheorem{proposition}[theorem]{Proposition}

\theoremstyle{definition}
\newtheorem{definition}{Definition}[section]
\newtheorem{example}[definition]{Example}

\theoremstyle{remark}
\newtheorem{remark}{Remark}[section]

\newcommand{\orcid}[1]{\href{https://orcid.org/#1}{\includegraphics[width=8pt]{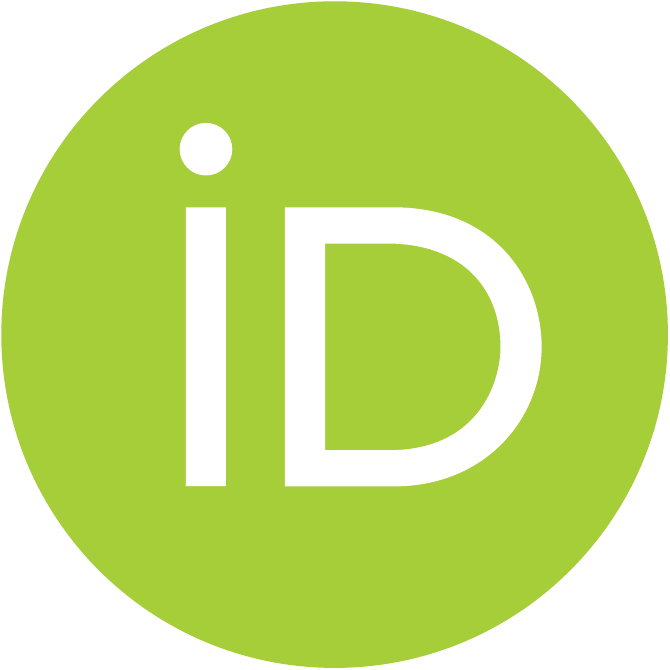}}}

\newcommand\nnfootnote[1]{%
  \begin{NoHyper}
  \renewcommand\thefootnote{}\footnote{#1}%
  \addtocounter{footnote}{-1}%
  \end{NoHyper}
}

\title{
A synopsis of the non-invertible, two-dimensional, border-collision normal form with applications to power converters.
}
\author{
H.O.~Fatoyinbo$^\dagger$  and
D.J.W.~Simpson$^\ddagger$ \\\\
$^\dagger$EpiCentre, School of Veterinary Science\\
$^\ddagger$School of Mathematical and Computational Sciences\\
Massey University\\
Palmerston North, 4410\\
New Zealand
}
\maketitle

% keywords: bifurcation, piecewise-smooth, power converter
% MSC codes:
%		39A28 -- Bifurcation theory
%		34A38 -- Hybrid Systems

\begin{abstract}

The border-collision normal form is a canonical form for two-dimensional, continuous maps comprised of two affine pieces. In this paper we provide a guide to the dynamics of this family of maps in the non-invertible case where the two pieces fold onto the same half-plane. We identify parameter regimes for the occurrence of key bifurcation structures, such as period-incrementing, period-adding, and robust chaos. We then apply the results to a classic model of a boost converter for adjusting the voltage of direct current. It is known that for one combination of circuit parameters the model exhibits a border-collision bifurcation that mimics supercritical period-doubling and is non-invertible due to the switching mechanism of the converter. We find that over a wide range of parameter values, even though the dynamics created in border-collision bifurcations is in general extremely diverse, the bifurcation in the boost converter can only mimic period-doubling, although it can be subcritical.

\end{abstract}

\nnfootnote{E-mail addresses: H.O. Fatoyinbo (\url{h.fatoyinbo@massey.ac.nz}), D.J.W. Simpson (\url{d.j.w.simpson@massey.ac.nz})}

%===============================================================================
\section{Introduction}
\label{sec:intro}

Periodic and non-periodic oscillations in systems of ordinary differential equations
are usually analysed by constructing a return map.
In classical settings, where the differential equations induce a unique smooth flow,
such maps are typically smooth and invertible, at least locally \cite{Me07}.
However, for piecewise-smooth differential equations and hybrid systems,
return maps are commonly piecewise-smooth \cite{DiBu08}.

The phase space of a piecewise-smooth map is characterised by the presence of
one or more switching manifolds where the map is nonsmooth.
This nonsmoothness causes the dynamics to change in a fundamental way at border-collision bifurcations
where a fixed point collides with a switching manifold as parameters are varied.
Under quite general conditions, these dynamics are captured by a
piecewise-linear family known as the border-collision normal form \cite{Di03,NuYo92,Si16}.
In two dimensions this family can be written as
\begin{equation}
\begin{bmatrix} x \\ y \end{bmatrix} \mapsto
\begin{cases}
\begin{bmatrix} \tau_L x + y + \mu \\ -\delta_L x \end{bmatrix}, & x \le 0, \\
\begin{bmatrix} \tau_R x + y + \mu \\ -\delta_R x \end{bmatrix}, & x \ge 0,
\end{cases}
\label{eq:bcnf}
\end{equation}
where the line $x = 0$ is the {\em switching manifold},
the parameter $\mu \in \mathbb{R}$ controls the border-collision bifurcation,
and $\tau_L, \delta_L, \tau_R, \delta_R \in \mathbb{R}$ are additional parameters.

The normal form \eqref{eq:bcnf} is well-studied
as it arises in diverse applications, and, as an extension of the Lozi map \cite{Lo78},
serves as a minimal model for chaotic and highly nonlinear dynamics.
The dynamics of \eqref{eq:bcnf} is remarkably rich --- it exhibits chaos robustly \cite{BaYo98,GlSi21},
can have any number of coexisting attractors \cite{Si14},
of which all could be chaotic \cite{PuRo18,Si22c}.
Most studies of \eqref{eq:bcnf} have focussed on parameter regimes where \eqref{eq:bcnf} is invertible,
i.e.~$\delta_L \delta_R > 0$.
For the invertible, dissipative case, a review is provided by \cite{BaGr99};
other works that characterise parameter space in some detail include \cite{SiMe08b,Si20e,ZhMo06b,SuGa08}.

However, it is perhaps under-appreciated that
return maps of piecewise-smooth dynamical systems
are often, not only piecewise-smooth, but also non-invertible.
This is because a switch to a different mode of operation
can readily cause the flow to fold back onto itself,
as illustrated below for power converters.
Some analysis of \eqref{eq:bcnf} has been done in the non-invertible case.
For example period-adding is illustrated numerically in \cite{SuGa08},
global bifurcations of a chaotic attractor in an equivalent map are described in \cite{MiRa96},
and two-dimensional attractors are identified in \cite{Gl16e}.
In this paper we provide an overview of the dynamics of \eqref{eq:bcnf} when it is non-invertible,
specifically with $\delta_L \delta_R < 0$.
For the special case $\delta_L \delta_R = 0$,
where the long-term dynamics are essentially one-dimensional,
refer to \cite{Ko05,SzOs09}.

We begin in \S\ref{sec:bcbs} by showing how the piecewise-linear normal form \eqref{eq:bcnf}
applies to border-collision bifurcations of arbitrary, two-dimensional, piecewise-smooth maps.
We then study \eqref{eq:bcnf} subject to $\delta_R < 0 < \delta_L$,
the specific signs being chosen without loss of generality.
The dynamical complexity of \eqref{eq:bcnf} means we cannot hope to characterise all dynamics,
and for this reason our approach is to chart the essential features.
With $\mu < 0$ the dominate bifurcation structures are period-adding and robust chaos, \S\ref{sec:muNegative},
while with $\mu > 0$ the map exhibits period-incrementing and robust chaos, \S\ref{sec:muPositive}.

In \S\ref{sec:powerConverters} we illustrate the results with the power converter model of Deane \cite{De92c}.
This model exhibits a border-collision bifurcation, and
by determining the part of the parameter space of the normal form that
this bifurcation corresponds to, we can use our results to characterise the bifurcation.
Although in general the dynamics created in border-collision bifurcations is extremely diverse,
it appears this bifurcation acts exclusively as a piecewise-smooth version of period-doubling.
Finally \S\ref{sec:conc} provides concluding remarks.

%===============================================================================
\section{Border-collision bifurcations and the normal form}
\label{sec:bcbs}

A border-collision bifurcation occurs when a fixed point of a piecewise-smooth map collides with a switching manifold.
Here we consider a two-dimensional map with variables $u, v \in \mathbb{R}$ and parameter $\eta \in \mathbb{R}$.
We assume a border-collision bifurcation occurs at the origin $(u,v) = (0,0)$ when $\eta = 0$, and
we wish to understand the dynamics in a neighbourhood of $(u,v;\eta) = (0,0;0)$.

We assume the switching manifold is smooth, at least locally,
so there exists a smooth coordinate change
%valid in a neighbourhood of $(u,v;\eta) = (0,0;0)$,
that shifts the switching manifold to the line $u=0$ \cite{DiBu01}.
Then, assuming the map is continuous and piecewise-$C^2$, locally it has the form
\begin{equation}
\begin{bmatrix} u \\ v \end{bmatrix} \mapsto \begin{cases}
\begin{bmatrix}
a^L_{11} u + a_{12} v + b_1 \eta \\
a^L_{21} u + a_{22} v + b_2 \eta
\end{bmatrix}
+ \cO \left( \left( |u| + |v| + |\eta| \right)^2 \right), & u \le 0, \\
\begin{bmatrix}
a^R_{11} u + a_{12} v + b_1 \eta \\
a^R_{21} u + a_{22} v + b_2 \eta
\end{bmatrix}
+ \cO \left( \left( |u| + |v| + |\eta| \right)^2 \right), & u \ge 0,
\end{cases}
\label{eq:fGeneral}
\end{equation}
%for some values of the eight coefficients $a^L_{11}, a_{12}, \ldots, b_2 \in \mathbb{R}$.
for some $a^L_{11}, a^L_{21}, a^R_{11}, a^R_{21}, a_{12}, a_{22}, b_1, b_2 \in \mathbb{R}$.
Note that the $v$ and $\eta$ coefficients of the two pieces of \eqref{eq:fGeneral} are the same.
This is a consequence of the assumed continuity of \eqref{eq:fGeneral} on $u = 0$.
Next we work to bring \eqref{eq:fGeneral} into the normal form \eqref{eq:bcnf}.

%-------------------------------------------------------------------------------
\subsection{A derivation of the normal form}

By construction the origin is a fixed point of \eqref{eq:fGeneral} when $\eta = 0$.
We now consider fixed points of \eqref{eq:fGeneral} for small $\eta \in \mathbb{R}$.
Let
\begin{equation}
c = \left( 1 - a_{22} \right) b_1 + a_{12} b_2 \,.
\label{eq:xi}
\end{equation}
Then the left ($u \le 0$) piece of \eqref{eq:fGeneral} has the fixed point
\begin{equation}
\begin{bmatrix} u^L(\eta) \\ v^L(\eta) \end{bmatrix} =
\frac{1}{\left( 1 - a^L_{11} \right) \left( 1 - a_{22} \right) - a_{12} a^L_{21}}
\begin{bmatrix} c \\ a^L_{21} b_1 + \left( 1 - a^L_{11} \right) b_2 \end{bmatrix} \eta + \cO \left( \eta^2 \right),
\label{eq:xLyL}
\end{equation}
assuming the denominator in \eqref{eq:xLyL} is non-zero.
Similarly the right ($u \ge 0$) piece of \eqref{eq:fGeneral} has the fixed point
\begin{equation}
\begin{bmatrix} u^R(\eta) \\ v^R(\eta) \end{bmatrix} =
\frac{1}{\left( 1 - a^R_{11} \right) \left( 1 - a_{22} \right) - a_{12} a^R_{21}}
\begin{bmatrix} c \\ a^R_{21} b_1 + \left( 1 - a^R_{11} \right) b_2 \end{bmatrix} \eta + \cO \left( \eta^2 \right),
\label{eq:xRyR}
\end{equation}
assuming its denominator is non-zero.
In a sufficiently small neighbourhood of $(u,v;\eta) = (0,0;0)$,
these are the only fixed points of \eqref{eq:fGeneral}.

Notice we require $c \ne 0$ for the fixed points to move away from the switching manifold
at a rate that is asymptotically proportional to $\eta$.
Thus $c \ne 0$ is the {\em transversality condition} \cite{GuHo86,Ku04}
that ensures $\eta$ unfolds the border-collision bifurcation in a generic fashion.

In view of the switching condition in \eqref{eq:fGeneral},
$\left( u^L, v^L \right)$ is a fixed point of \eqref{eq:fGeneral} only if $u^L \le 0$,
in which case we say it is {\em admissible}.
Similarly $\left( u^R, v^R \right)$ is an admissible fixed point of \eqref{eq:fGeneral} if $u^R \ge 0$.
If $u^L < 0$, the stability of $\left( u^L, v^L \right)$ is governed by the eigenvalues
of the Jacobian matrix of \eqref{eq:fGeneral} evaluated at $\left( u^L, v^L \right)$.
This matrix is $\begin{bmatrix} a^L_{11} & a_{12} \\ a^L_{21} & a_{22} \end{bmatrix} + \cO(\eta)$.
To leading order its trace and determinant are
\begin{align}
\tau_L &= a^L_{11} + a_{22} \,, &
\delta_L &= a^L_{11} a_{22} - a_{12} a^L_{21} \,.
\label{eq:tauLdeltaL}
\end{align}
%Similarly, for the right piece of \eqref{eq:fGeneral},
%the trace and determinant of the Jacobian matrix are, to leading order,
We similarly define
\begin{align}
\tau_R &= a^R_{11} + a_{22} \,, &
\delta_R &= a^R_{11} a_{22} - a_{12} a^R_{21} \,,
\label{eq:tauRdeltaR}
\end{align}
for the right piece of \eqref{eq:fGeneral}.
By removing the nonlinear terms from each piece of \eqref{eq:fGeneral}
and applying the coordinate change
%Now remove the nonlinear terms from each piece of \eqref{eq:fGeneral}.
%This results in a piecewise-linear map that, ostensibly, approximates the dynamics of \eqref{eq:fGeneral}
%for small values of $x$, $y$, and $\eta$.
%Moreover, the coordinate change
\begin{align}
x &= u, \\
y &= -a_{22} u + a_{12} v + \left( a_{22} b_1 - a_{12} b_2 \right) \eta, \\
\mu &= c \eta,
\label{eq:coordinateChange}
\end{align}
we arrive at the normal form \eqref{eq:bcnf}
with $\tau_L$, $\delta_L$, $\tau_R$, and $\delta_R$
given by \eqref{eq:tauLdeltaL} and \eqref{eq:tauRdeltaR}.
This coordinate change is invertible when $c \ne 0$ (discussed above)
and $a_{12} \ne 0$ (otherwise \eqref{eq:fGeneral} decouples into two one-dimensional maps).

%-------------------------------------------------------------------------------
%\subsection{Basic properties of the normal form}
\subsection{The utility of the normal form}

The dynamics of the normal form \eqref{eq:bcnf}
approximates the dynamics of \eqref{eq:fGeneral}
for small values of $u$, $v$, and $\eta$.
Presently there is little mathematical theory clarifying
the validity of this approximation,
but in practice the normal form is useful for characterising the dynamics created in
border-collision bifurcations.
For example, if the normal form exhibits a hyperbolic periodic solution,
this solution also exists for \eqref{eq:fGeneral} \cite{Si16}.
This type of persistence result
has recently been extended to chaotic attractors \cite{SiGl21}.

Since the normal form is piecewise-linear,
the structure of its dynamics is independent of the magnitude of $\mu$.
If $\mu < 0$ its value can be scaled to $-1$,
while if $\mu > 0$ its value can be scaled to $1$.
We therefore study the normal form with $\mu = -1$ to understand the dynamics
on one side of the border-collision bifurcation,
and with $\mu = 1$ to understand the dynamics on the other side of the border-collision bifurcation.

With $\mu = 0$ the origin is a fixed point of the normal form.
If it is asymptotically stable then \eqref{eq:fGeneral} has a local attractor
on each side of the border-collision bifurcation \cite{Si20b}.
However, the normal form is non-differentiable at the origin,
so its stability can be difficult to ascertain \cite{DoKi08,Ga92}.
For instance in the non-invertible case the origin can be asymptotically stable even when
both pieces of the map are area-expanding \cite{Si20d}.

If $\delta_L \delta_R < 0$, then, in view of the substitution $(x,y,\mu) \mapsto (-x,-y,-\mu)$
that leaves the normal form invariant other than switching `left' and `right',
we can assume $\delta_L > 0$ and $\delta_R < 0$.
With $\delta_L > 0$ and $\delta_R < 0$
the normal form maps both left and right half-planes to the upper half-plane ($y \ge 0$).
%In the terminology of Mira {\em et.~al.}~\cite{MiGa96},
%the image of the switching manifold (i.e.~the line $y=0$) is a {\em critical line} and \eqref{eq:bcnf} is of class $Z_0$-$Z_2$.

%-------------------------------------------------------------------------------
\subsection{An example of the dynamics of the normal form}

\begin{figure}
    \centering
    \includegraphics[scale=0.6]{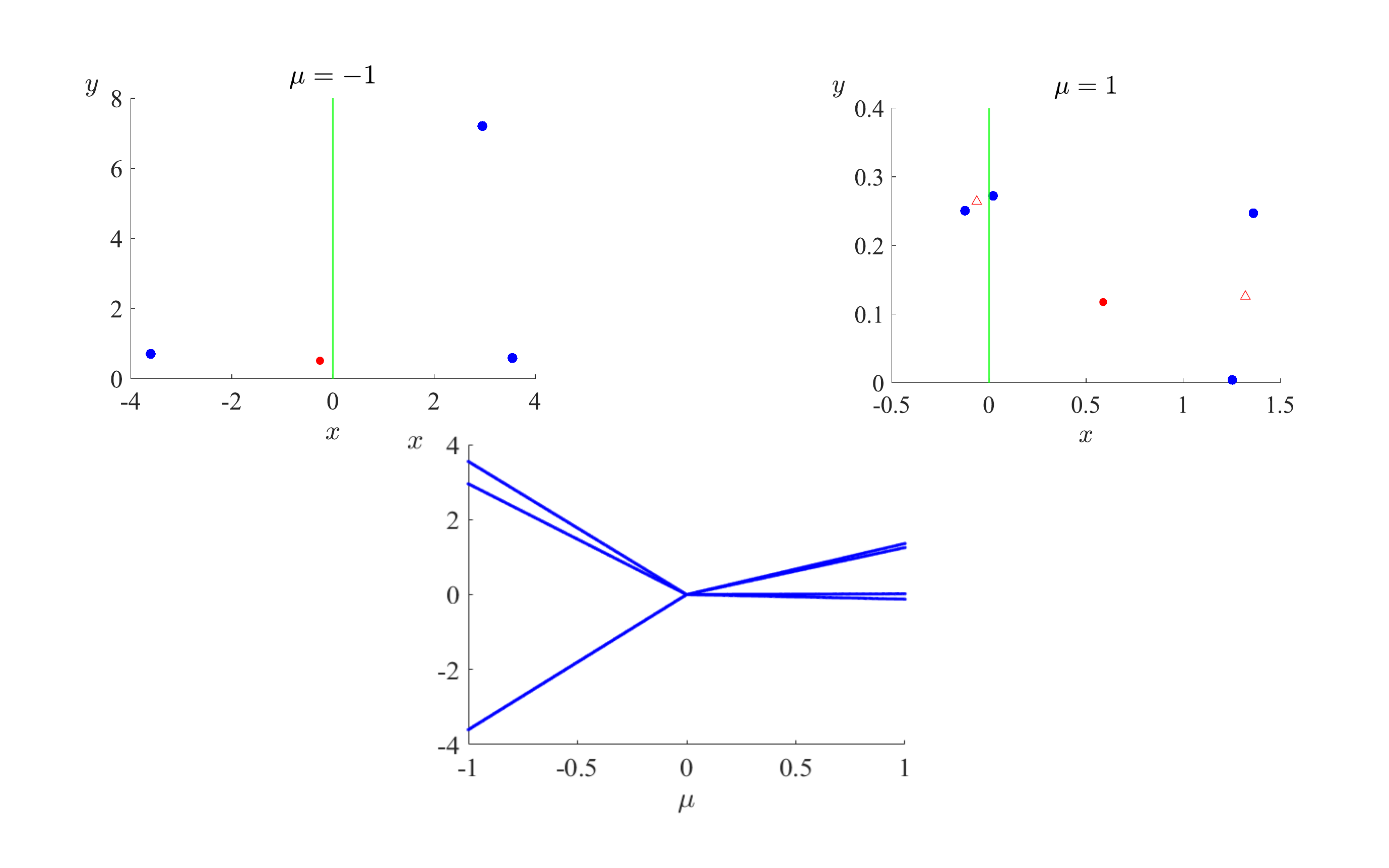}
    \caption{
   The lower plot is a bifurcation diagram of the normal form \eqref{eq:bcnf} with \eqref{eq:paramEx1}.
   As the value of $\mu$ is increased through $0$,
   a stable period-$3$ solution changes to a stable period-$4$ solution.
   The upper plots are phase portraits with $\mu = \pm 1$
   (red circle: unstable fixed point;
   red triangles: unstable period-$2$ solution;
   blue circles: stable period-$3$ and period-$4$ solutions;
   green line: switching manifold).}
 \label{fig:bif_3_to_4-pp}
\end{figure}

Fig.~\ref{fig:bif_3_to_4-pp}
shows a bifurcation diagram and phase portraits
of the normal form \eqref{eq:bcnf} with
\begin{equation}
(\tau_L,\delta_L,\tau_R,\delta_R) = (-1,2,-1,-0.2).
\label{eq:paramEx1}
\end{equation}
%with instead $\delta_R = -0.1$ the figure is too skewed and doesn't look nice
Here $\delta_L > 0$ and $\delta_R < 0$ so all invariant sets
lie in the upper half-plane ($y \ge 0$).
For this example the border-collision bifurcation brings about a transition
from a stable period-$3$ solution to a stable period-$4$ solution.
%A variety of similar transitions are illustrated in the original work of
%Nusse and Yorke \cite{NuYo92}.
%With $\mu < 0$ there exists a stable period-$3$ solution.
%This solution has a point in 
As with all bounded invariant sets of the normal form,
these solutions contract linearly to the origin as $\mu \to 0$,
and this is evident in the bifurcation diagram.
The period-$3$ solution for $\mu < 0$ consists of one point in the left half-plane
and two points in the right half-plane.
Its {\em symbolic representation} is therefore $LRR$ (or any cyclic permutation of this)
and we refer to it as an $LRR$-cycle.
Similarly the period-$4$ solution for $\mu > 0$ is an $LRRR$-cycle.
Based on symbolic representations,
we can characterise the existence, admissibility, and stability of
periodic solutions in a general manner \cite{Si16,SiMe10}.

%===============================================================================
\section{The dynamics for $\mu < 0$}
\label{sec:muNegative}

\begin{figure}[htbp]
\centering
\includegraphics[scale=0.65]{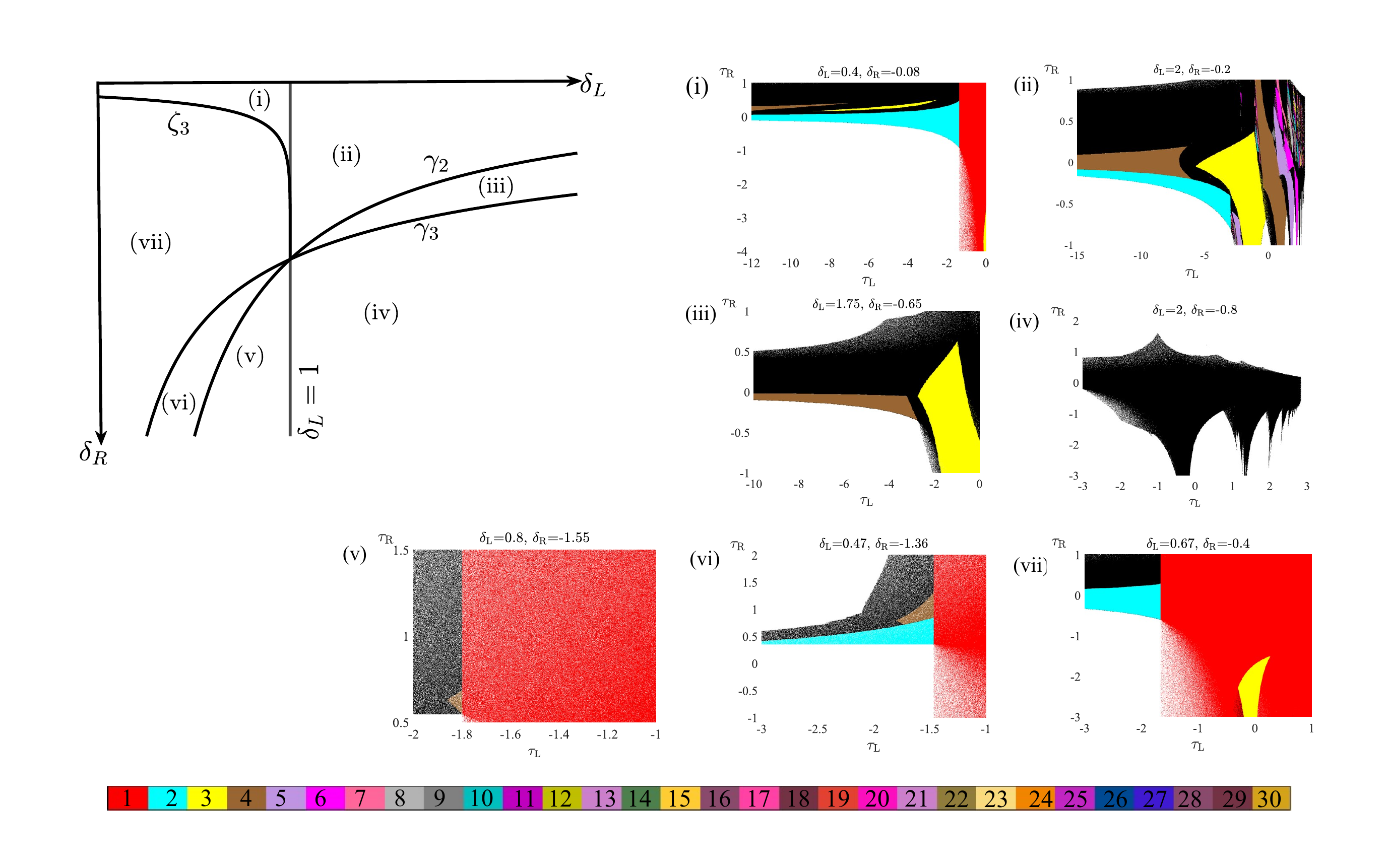}
\caption{
Seven representative two-parameter bifurcation diagrams of the normal form \eqref{eq:bcnf} with $\mu = -1$
obtained by fixing $\delta_L$ and $\delta_R$ and varying $\tau_L$ and $\tau_R$
(black: chaotic, quasi-periodic, or high period attractor;
white: no attractor;
red: stable fixed point;
cyan: stable period-$2$ solution;
yellow: stable period-$3$ solution;
other colours: periods $4$--$30$ as indicated in the colour bar).
The figure also shows how the fourth quadrant of the $(\delta_L,\delta_R)$-plane
can be divided into subsets where the bifurcation diagrams have similar features.
%The curves $\delta_L = \frac{-1}{\delta_R}$
%and $\delta_L = \frac{1}{\delta_R^2}$ are labelled $\gamma_2$ and $\gamma_3$ respectively.
}
\label{fig:combo1}
\end{figure}

Our aim here is to obtain a broad understanding of the attractors
of \eqref{eq:bcnf} with $\mu < 0$
and $\delta_L > 0$, $\delta_R < 0$, $\tau_L \in \mathbb{R}$, and $\tau_R \in \mathbb{R}$.
As explained above, by scaling it is sufficient to consider $\mu = -1$.
We have found it helpful to study two-dimensional slices of parameter space
defined by fixing the values of $\delta_L$ and $\delta_R$,
and study how the dynamics on these slices differs for different values of $\delta_L$ and $\delta_R$.
This is summarised by Fig.~\ref{fig:combo1}.
In the top left this figure shows the open fourth quadrant
\begin{equation}
Q_4 = \left\{ (\delta_L,\delta_R) \middle| \delta_L > 0, \delta_R < 0 \right\},
\label{eq:Q4}
\end{equation}
and has four curves that divide $Q_4$ into seven subsets
where $(\tau_L,\tau_R)$-two-parameter bifurcation diagrams have similar features.
This division is based on the existence of period-$n$ periodicity regions for $n = 1,2,3$.
Certainly one could divide $Q_4$ further by considering more features in the bifurcation diagrams,
but it is not clear that this would be helpful \cite{Gl17b}.

%-------------------------------------------------------------------------------
\subsection{Numerical methods}

Before we describe the seven bifurcation diagrams in
Fig.~\ref{fig:combo1} in more detail, we first explain how they were computed.
For each point in a $1000 \times 1000$ equispaced grid of $(\tau_L,\tau_R)$ values,
we computed $N = 10^4$ iterates $(x_i,y_i)$ of the forward orbit of a random initial point $(x_0,y_0)$.
If the norm of $(x_i,y_i)$ exceeded $10^{6}$ for any $i$
we concluded the orbit diverges and coloured this parameter point white.
Otherwise we looked to see if the orbit converges to a periodic solution.
If we could find a smallest $1 \le p \le 30$
for which the norm of $(x_N,y_N) - (x_{N-p},y_{N-p})$ was less than a tolerance of $10^{-6}$,
we concluded the orbit converges to a period-$p$ solution
and coloured the point according to the colour bar.
If no such period was detected the point was coloured black.
In this case the orbit most likely converges to either a periodic solution with period greater than $30$,
a quasi-periodic attractor, or a chaotic attractor.
Additional techniques were used to compute the period-$3$ regions in (i) and (vii).
Following \cite{Si16} for these we solved for the $LRR$-cycle exactly
and checked admissibility and stability explicitly.

By using random initial points the bifurcation diagrams
reveal areas where multiple attractors coexist
and areas where attractors are not globally stable.
For example the bottom three bifurcation diagrams
are speckled white because the attractors are not globally attracting
so some initial points give forward orbits that diverge.

%-------------------------------------------------------------------------------
\subsection{Seven different two-parameter bifurcation diagrams}

First consider $(\delta_L,\delta_R)$ in subset (i).
In this case any $(\tau_L,\tau_R)$-bifurcation diagram
has a vertical red strip, $-\delta_L - 1 < \tau_L < \delta_L + 1$,
where the left half-map has a stable fixed point. %$(x^L,y^L)$.
As we cross the $\tau_L = -\delta_L - 1$ boundary of this strip, from right to left,
this fixed point attains an eigenvalue of $-1$ and becomes unstable.
This is not a period-doubling bifurcation as the normal form does not have
the required nonlinearity,
nevertheless a stable period-$2$ solution ($LR$-cycle) exists
in a certain region (coloured cyan).
This region is bounded below by the curve
$\tau_R =\frac{(1+\delta_{L})(1+\delta_{R})}{\tau_{L}}$, where
the $LR$-cycle has an eigenvalue of $-1$,
and bounded above by $\tau_R = \frac{(\delta_{L}-1)(1-\delta_{R})}{\tau_{L}}$,
where it has an eigenvalue of $1$.
Fig.~\ref{fig:combo1}(i) also has regions where
there exists a stable $LRR$-cycle (yellow) and a stable $LRRR$-cycle (brown).
For different values of $(\delta_L,\delta_R)$ in subset (i),
our numerical explorations have found that other periodicity regions
can exist, particularly if $|\delta_R|$ is small,
but the corresponding symbolic itinerary always contains exactly one $L$.
%These are the only stable periodic solutions that seem to be possible in subset (i).

\begin{figure}[htbp]
\centering
\begin{subfigure}[b]{.5\linewidth}
    \centering
    \caption{}
    \includegraphics[width=.99\textwidth]{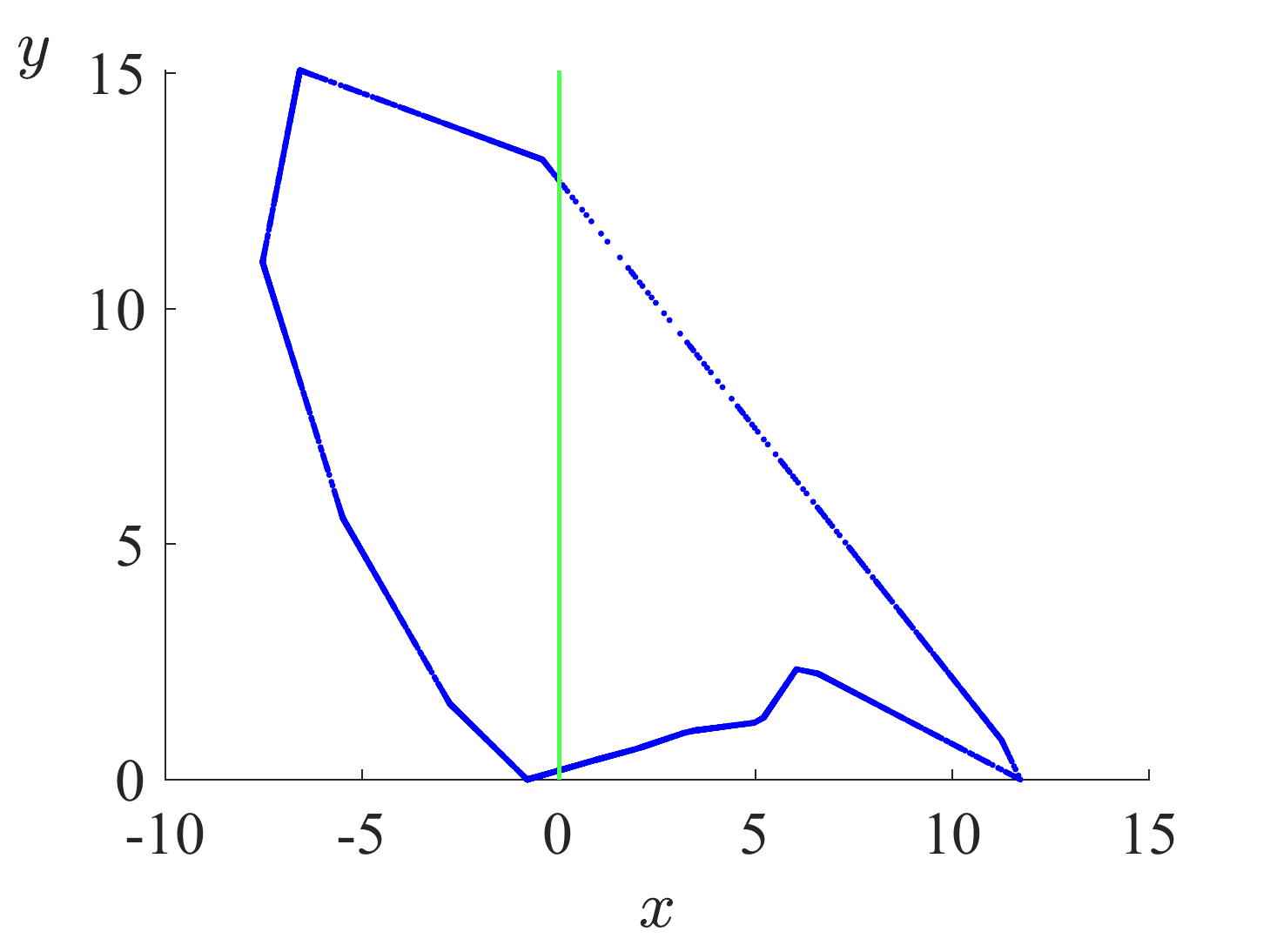}
  \end{subfigure}%
  \begin{subfigure}[b]{.5\linewidth}
    \centering
    \caption{}
    \includegraphics[width=.99\textwidth]{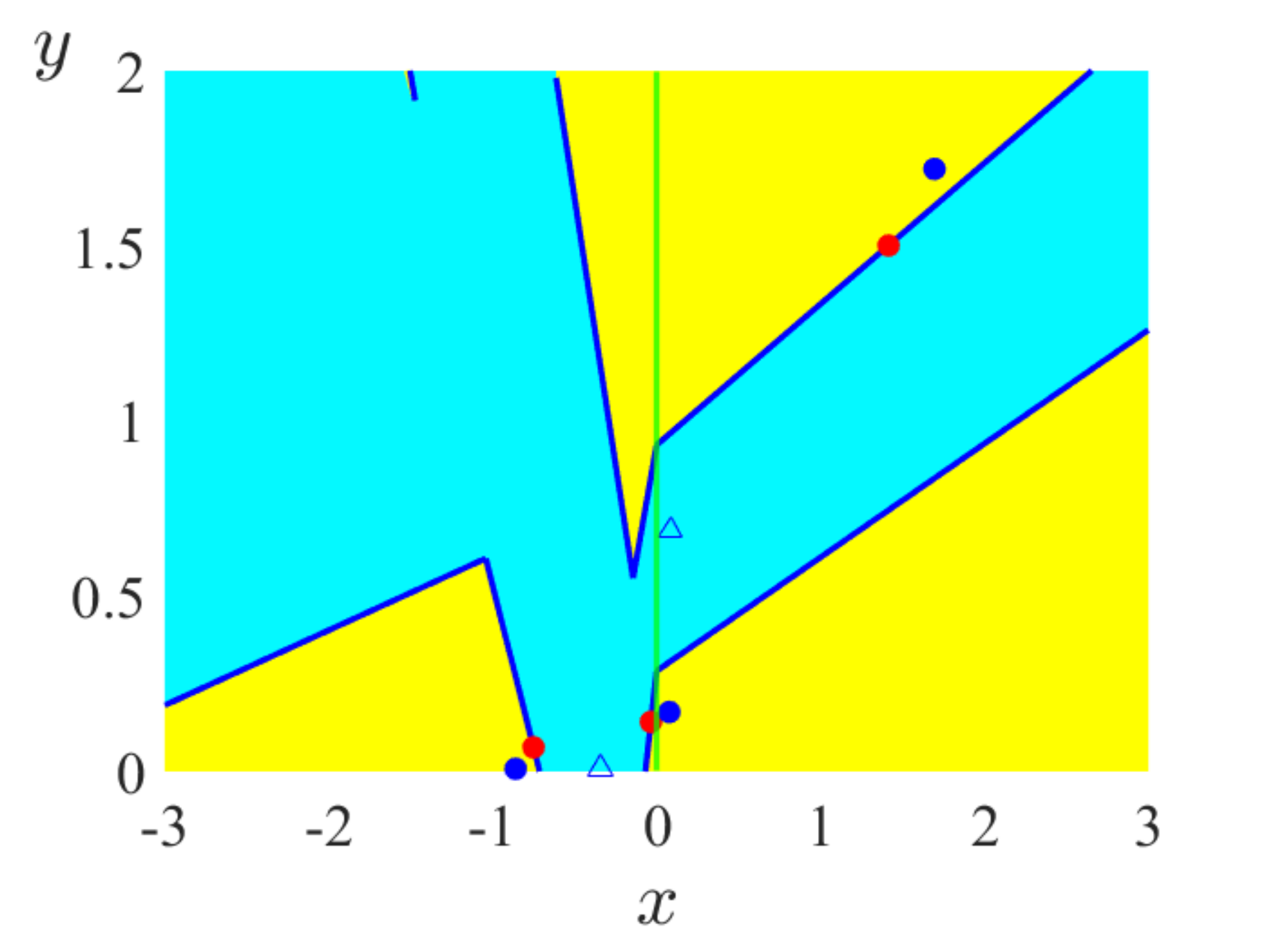}
  \end{subfigure}%  
  \caption{
  Panel (a) is a phase portrait of \eqref{eq:bcnf} with $\mu = -1$
  and $(\tau_L,\delta_L,\tau_R,\delta_R) = (2.2,2,0.6,-0.2)$
  showing an attracting invariant circle.
  Panel (b) is with instead
  $(\tau_L,\delta_L,\tau_R,\delta_R) = (-3.126,2,-0.3785,-0.1)$
  where the map has a stable $LR$-cycle (blue triangles)
  and a stable $LRR$-cycle (blue circles).
  The boundary between the basins of attraction
  of these solutions is the stable manifold of a saddle-type $LLR$-cycle (red circles).
  }
  \label{fig:negativePhasePortraits1}
\end{figure}

To move from subset (i) to (ii) %(in the central plot of Fig.~\ref{fig:combo1})
we cross the line $\delta_L = 1$
at which the vertical red strip vanishes.
On the line $\delta_L = 1$, and with $-2 < \tau_L < 2$,
the left half-map has purely imaginary eigenvalues.
For generic families of smooth maps this signals the occurrence of a Neimark-Sacker bifurcation
at which an invariant circle is created.
From a curve of Neimark-Sacker bifurcations
there emanate {\em Arnold tongues} where the dynamics on the invariant circle
includes a stable periodic solution (i.e.~is {\em mode-locked}) \cite{Ku04,Ar88}.
In our piecewise-linear setting, similar features arise \cite{SiMe08b}.
An invariant circle is often created, as in Fig.~\ref{fig:negativePhasePortraits1}(a).
However, since the map is non-invertible the interior of the circle
does not map to itself.

Fig.~\ref{fig:combo1}(ii) shows a typical bifurcation diagram for subset (ii).
Stable periodic solutions exist in the coloured regions (Arnold tongues)
which overlap in places resulting in coexisting attractors,
Fig.~\ref{fig:negativePhasePortraits1}(b) shows an example.
Overall the Arnold tongues have a period-adding structure as discussed below.

%Here we continue to describe the division of $Q_4$.
To move from subset (ii) to (iii)
we cross the curve $\delta_L = \frac{-1}{\delta_R}$, labelled $\gamma_2$,
which is where the cyan region vanishes.
This can be explained as follows.
The eigenvalues that govern the stability of an admissible $LR$-cycle
are those of the matrix $A_R A_L$ \cite{Si16}.
The curve $\gamma_2$ is where the determinant of this matrix is $-1$.
Below $\gamma_2$ we have $\det(A_R A_L) < -1$ so the $LR$-cycle
cannot be stable for any values of $\tau_L$ and $\tau_R$.

By next crossing the curve $\delta_L = \frac{1}{\delta_R^2}$, labelled $\gamma_3$,
the yellow region (for the existence of a stable $LRR$-cycle) vanishes for a similar reason.
One could further identify additional curves for higher periods
but this is complicated by the fact that for higher periods multiple symbolic representations occur.

\begin{figure}[htbp]
\centering
\begin{subfigure}[b]{.5\linewidth}
    \centering
    \caption{}
    \includegraphics[width=.99\textwidth]{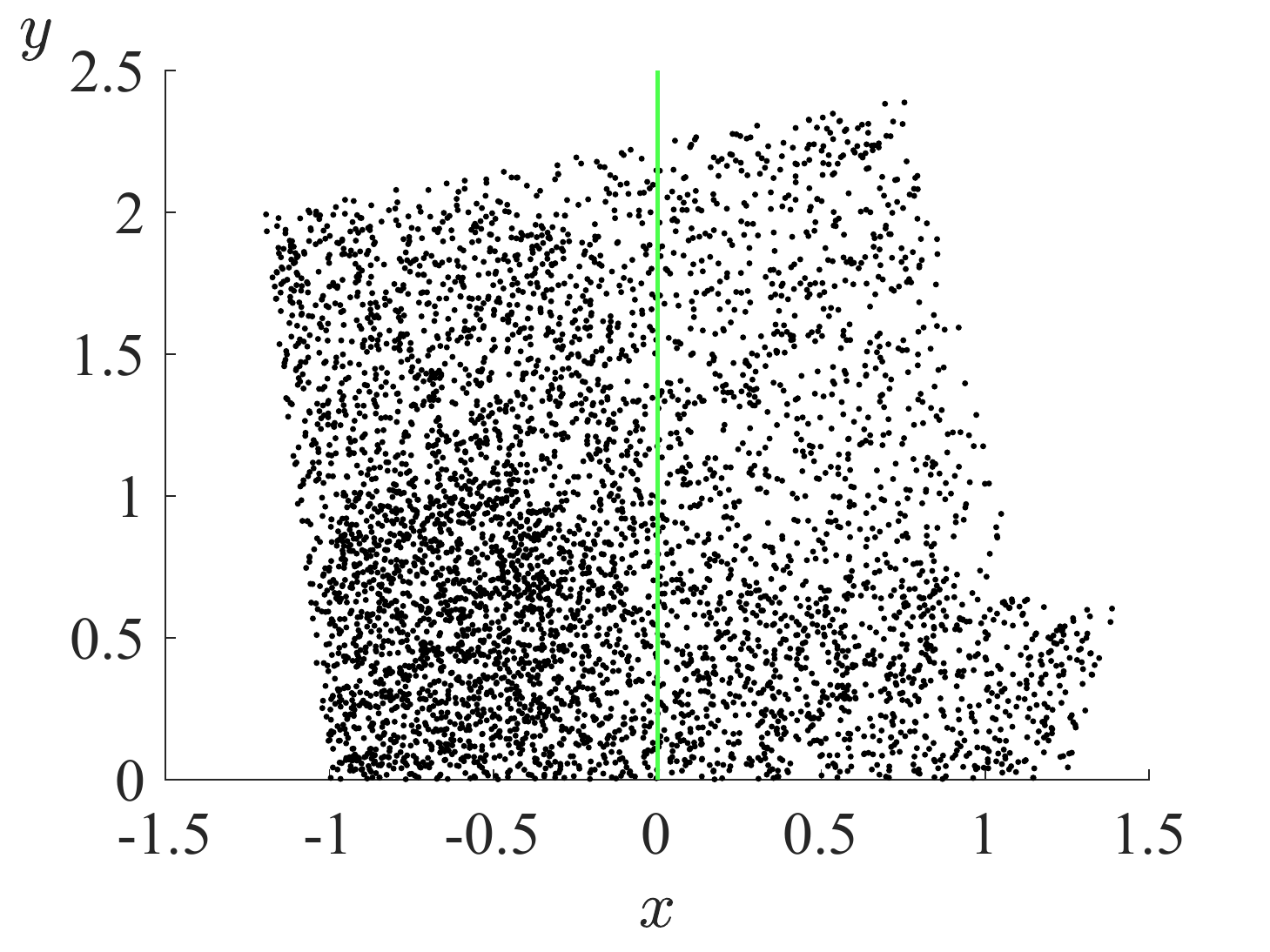}
    \label{fig:bcnf_bif}
  \end{subfigure}%
  \begin{subfigure}[b]{.5\linewidth}
    \centering
    \caption{}
    \includegraphics[width=.99\textwidth]{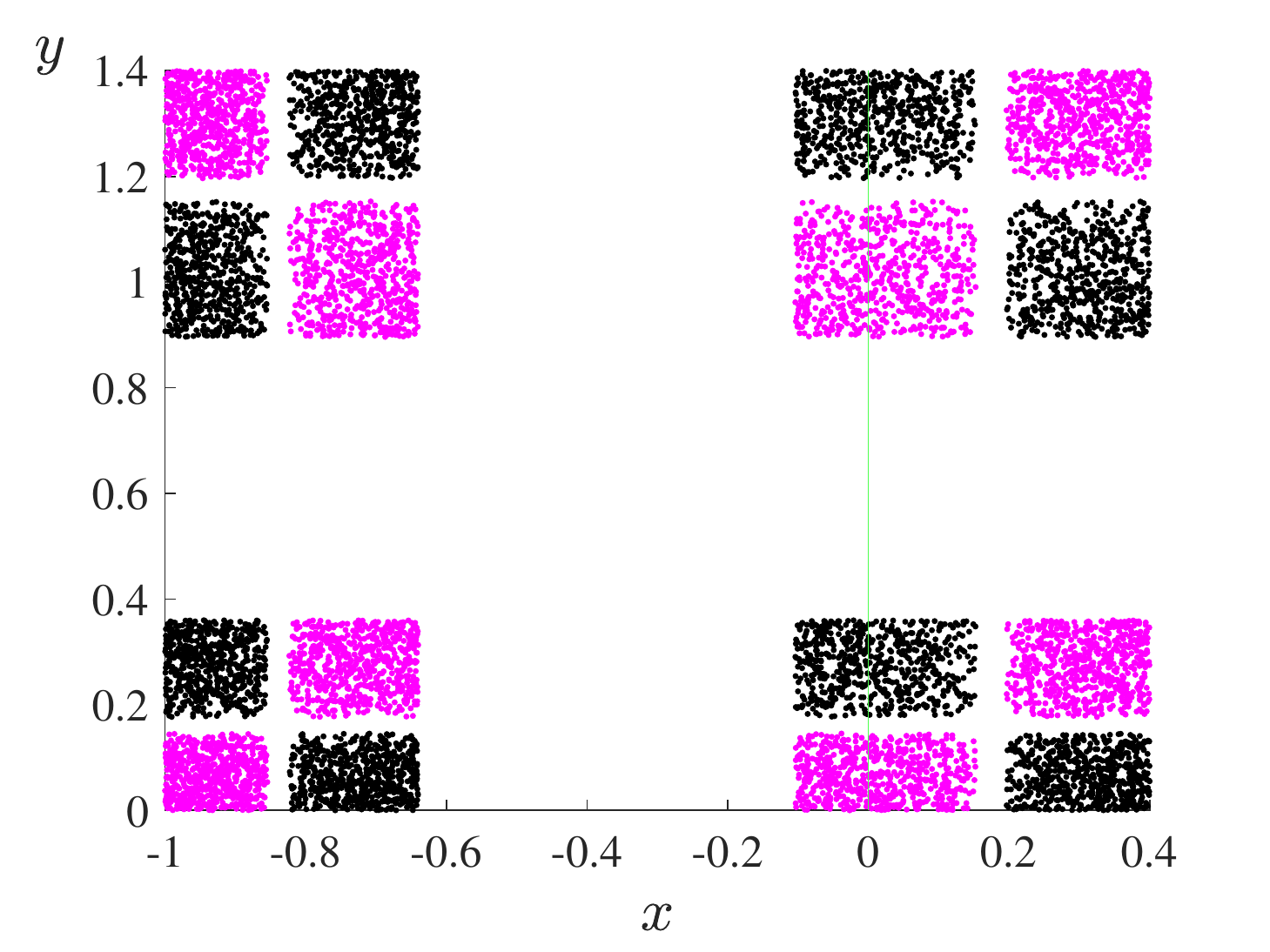}
    \label{fig:bcnf_PP}
  \end{subfigure}%  
  \caption{
  Panel (a) is a phase portrait of \eqref{eq:bcnf} with $\mu = -1$
  and $(\tau_L,\delta_L,\tau_R,\delta_R) = (0.2,2,0,-0.8)$
  showing a two-dimensional attractor, see \cite{Gl16e}.
  Panel (b) is with instead $(\tau_L,\delta_L,\tau_R,\delta_R) = (0,1.4,0,-0.9)$
  where there are two coexisting chaotic attractors, see \cite{PuRo18}.
  \label{fig:negativePhasePortraits2}
  }
\end{figure}

Our numerical explorations suggest that throughout subset (iv)
the only attractors that are possible are chaotic.
But unlike when the normal form is invertible
--- where chaotic attractors
are typically quasi-one-dimensional, formed from one or more one-dimensional unstable manifolds \cite{GlSi21}
--- in the non-invertible setting attractors can be two-dimensional, Fig.~\ref{fig:negativePhasePortraits2}(a).
Glendinning \cite{Gl16e} proved that throughout an open subset of parameter space
the normal form has a two-dimensional attractor
for one sign of $\mu$ and a stable fixed point for the other sign of $\mu$.
This shows that the sudden and local transition of a stable fixed point to a two-dimensional attractor
can occur via a border-collision bifurcation in a generic (codimension-one) fashion.
Coexisting chaotic attractors are also possible, Fig.~\ref{fig:negativePhasePortraits2}(b).
In fact for all $n \ge 1$ there exist parameter combinations
for which the normal form has $n$ coexisting chaotic attractors \cite{Si22c}.

Next as we move in $Q_4$ from (iv) into (v), then into (vi), and lastly into (vii),
we again cross the curves $\delta_L = 1$, $\gamma_2$, and $\gamma_3$, in this order.
Consequently the vertical red strip reappears (fixed point),
then the cyan region reappears ($LR$-cycle),
and lastly a yellow region reappears ($LRR$-cycle).
The difference between subsets (i) and (vii)
is that the yellow region has two connected components in subset (i),
but only one connected component in subset (vii).
The boundary between these subsets is a curve $\zeta_3$ derived in Appendix \ref{sec:A1}.

%-------------------------------------------------------------------------------
\subsection{Period-adding}

\begin{figure}[htbp]
\centering
\begin{subfigure}[b]{.5\linewidth}
    \centering
    \caption{}
     \includegraphics[width=.99\textwidth]{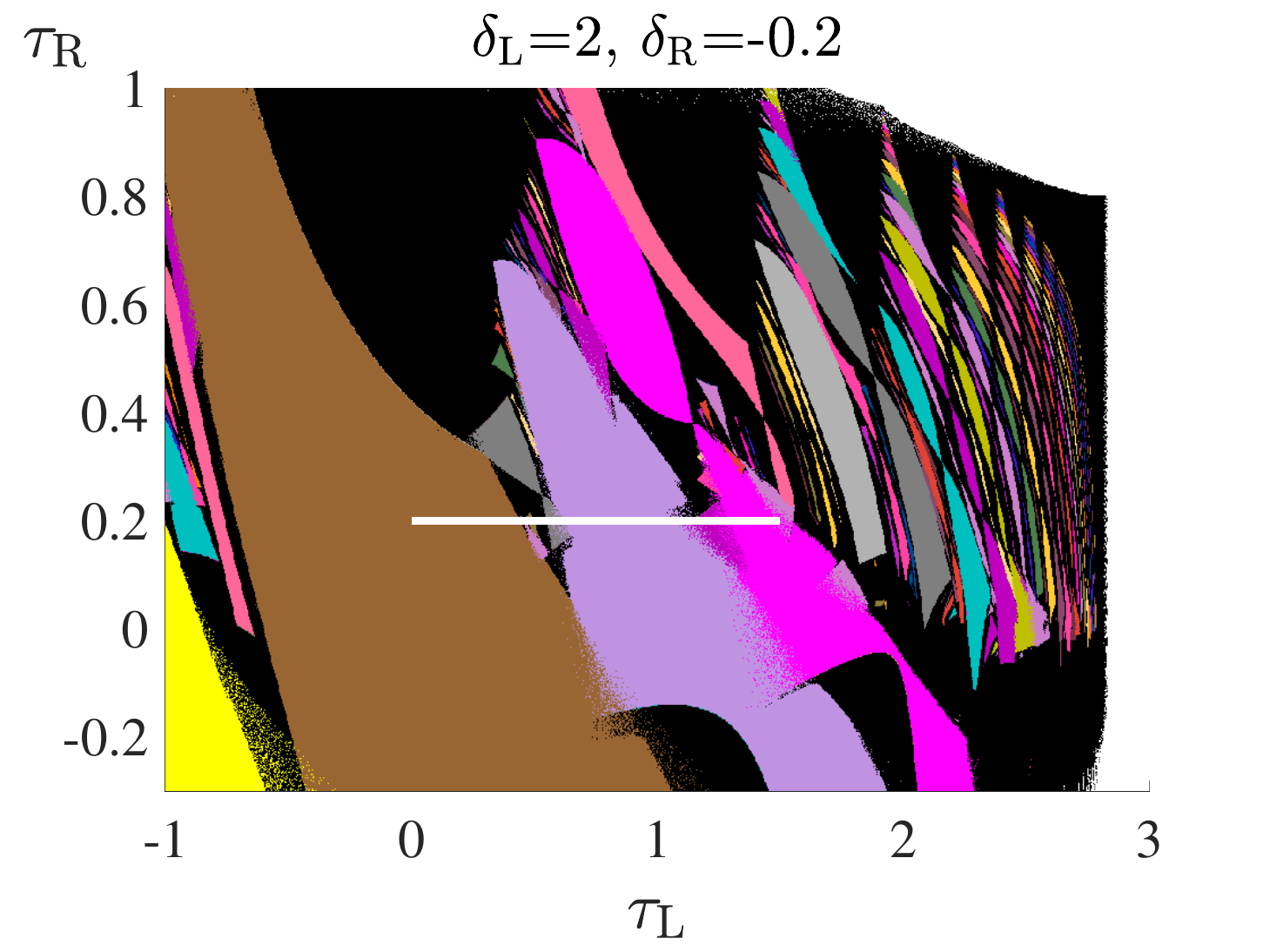}
    %\label{fig:zoomII}
  \end{subfigure}%
  \begin{subfigure}[b]{.5\linewidth}
    \centering
    \caption{}
    \includegraphics[width=.99\textwidth]{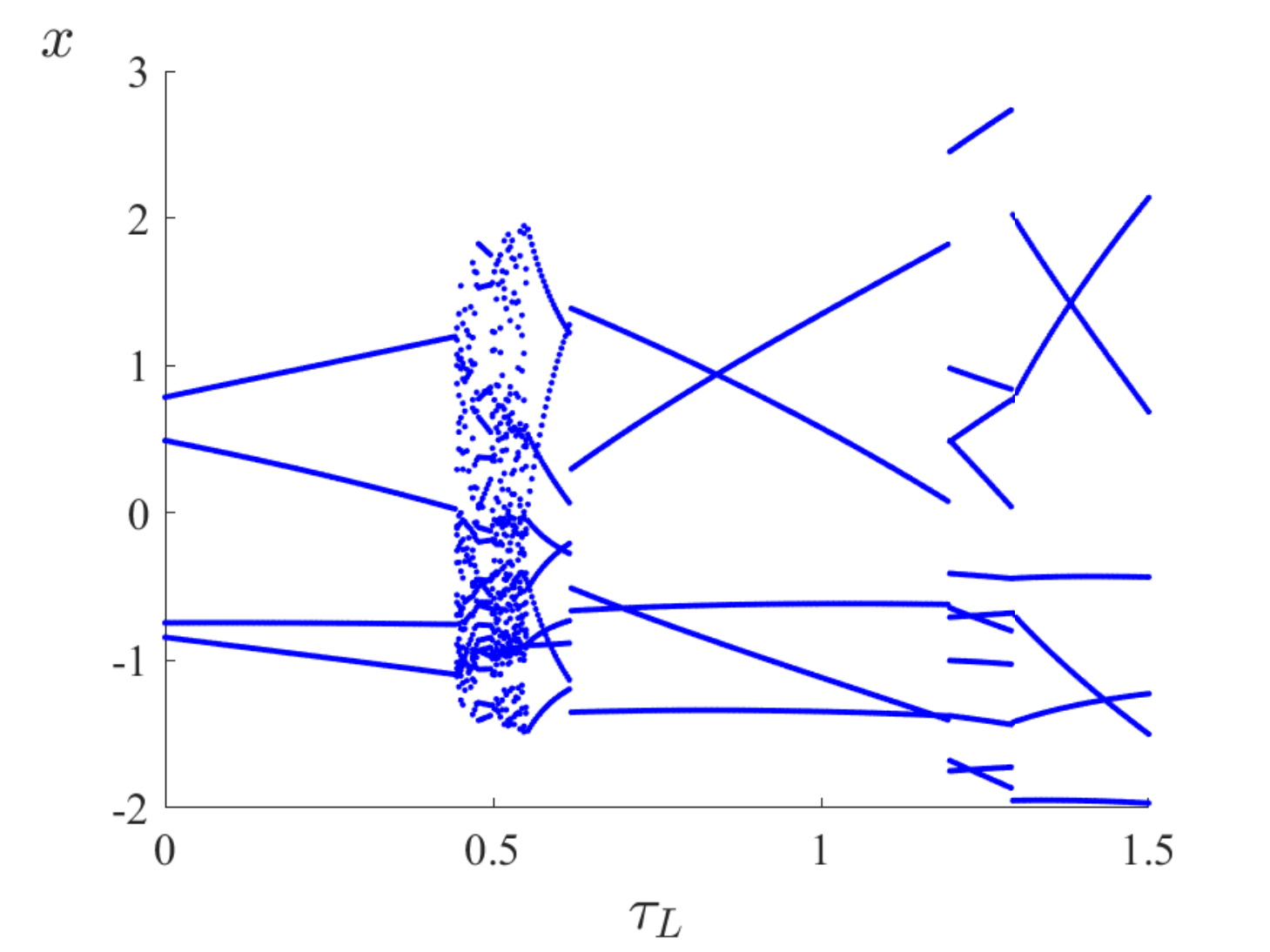}
    %\label{fig:periodadding}
  \end{subfigure}%  
  \caption{
  Panel (a) is a magnification of the two-dimensional bifurcation diagram, Fig.~\ref{fig:combo1}(ii),
  showing Arnold tongues (regions where periodic solutions are stable) up to period $30$.
  Panel (b) is a one-parameter bifurcation diagram obtained by further fixing $\tau_R = 0.2$.
  }
  \label{fig:negativePhasePortraits3}
\end{figure}

Arnold tongues are shown in Fig.~\ref{fig:combo1}(ii)
and more closely in the magnification, Fig.~\ref{fig:negativePhasePortraits3}(a).
While the precise location and shape of the Arnold tongues
depends on the values of $\delta_L$ and $\delta_R$,
they have a broadly consistent two-dimensional structure.
This structure is described in \cite{SuGa08} for our non-invertible setting,
in \cite{SzOs09} for the special case $\delta_L \delta_R = 0$,
and in several papers \cite{ZhYa10,GaGa03,LaMo06,Ti02}
(each motivated by a different physical application)
for the invertible case.

In the Arnold tongues only certain `rotational' symbolic itineraries are possible \cite{SiMe09}.
As in the classical smooth setting \cite{Fr90,Le01},
the periodic solution can be assigned a {\em rotation number}
$\frac{m}{p} \in (0,1)$, where $p$ is the period
and $m$ represents the number of rotations per period.
As we move left to right across Fig.~\ref{fig:negativePhasePortraits3}(a),
the rotation number decreases roughly monotonically.

Between any two Arnold tongues of frequencies $\frac{m_1}{p_1}$ and $\frac{m_2}{p_2}$
with $m_1 p_2 - m_2 p_1 = \pm 1$, there are usually more Arnold tongues, the largest of which
has frequency $\frac{m_1 + m_2}{p_1 + p_2}$.
Such {\em Farey addition} \cite{HaZh18,HaWr08} repeats indefinitely and, as in the smooth setting,
explains the ordering and relative size of the Arnold tongues.
One-parameter slices that cut horizontally
through Fig.~\ref{fig:negativePhasePortraits3}(a) consequently
display periodic windows where the period of a window
is often given by adding the periods of the largest windows on either side.
Fig.~\ref{fig:negativePhasePortraits3}(b) shows a typical example
where we can see period-$5$ and period-$6$ windows, and between them a period-$11$ window
(centred at about $\tau_L = 1.2$).
This structure is referred to as {\em period-adding} \cite{DiBu08}.

However, unlike in the smooth setting,
the Arnold tongues also exhibit structure in a second direction.
As we move up Fig.~\ref{fig:negativePhasePortraits3}(a)
the fraction of $L$'s in the symbolic itineraries decreases.
As we cross any {\em shrinking point}, where an Arnold tongue has
zero width, one $L$ in the symbolic itinerary associated with the stable periodic solution
changes to an $R$ \cite{SiMe09}.

%===============================================================================
\section{An overview of the dynamics with $\mu > 0$}
\label{sec:muPositive}

We now consider $\mu > 0$.
As discussed in \S\ref{sec:bcbs} it suffices to consider $\mu = 1$.

\begin{figure}[htbp]
\centering
\includegraphics[scale=0.6]{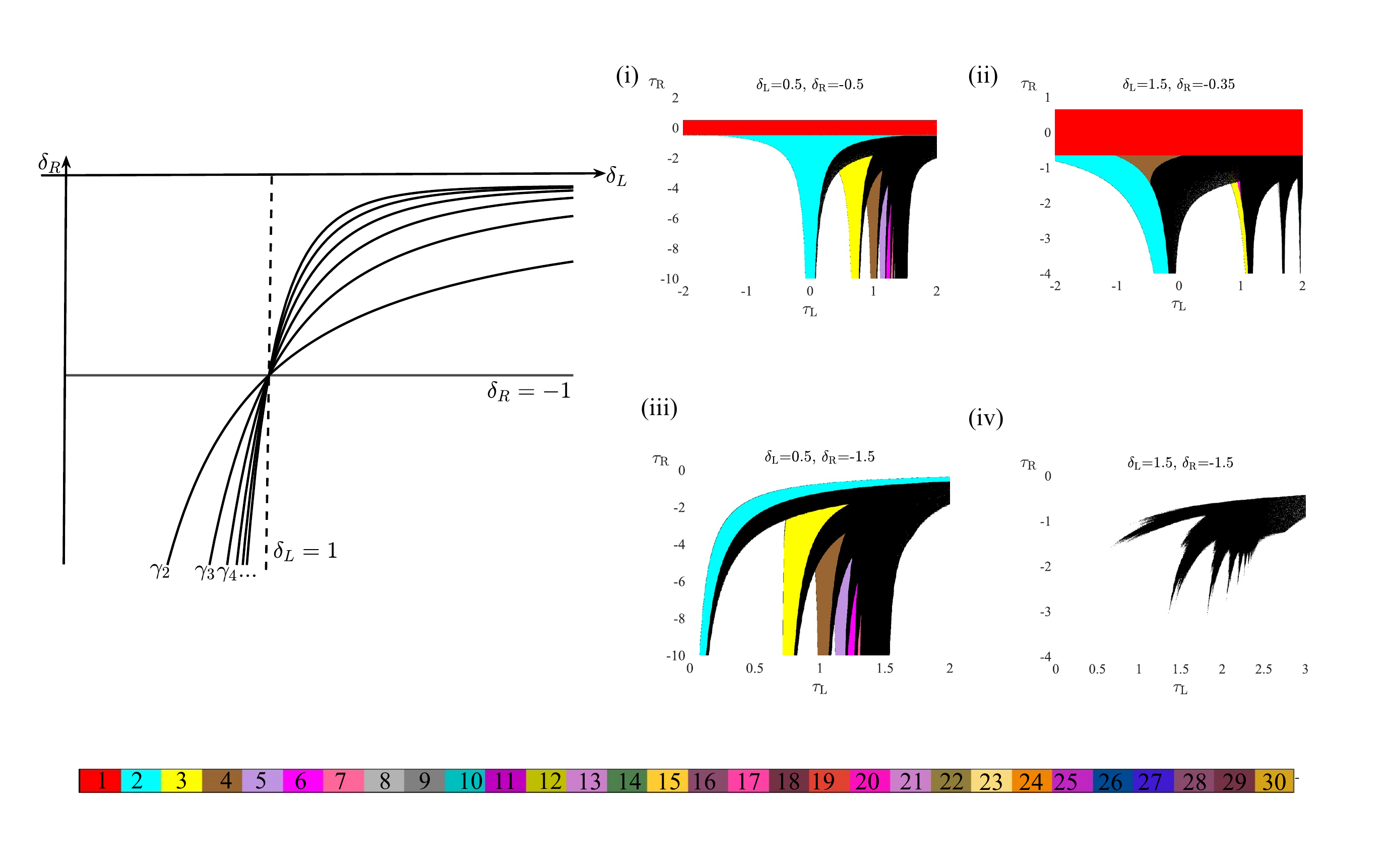}
\caption{
Four representative two-parameter bifurcation diagrams of \eqref{eq:bcnf}
with $\mu = 1$ using the same colour scheme as Fig.~\ref{fig:combo1}.
The left plot shows curve $\gamma_p$ that form a boundary for the existence of
a region in the bifurcation diagrams where the map has a stable $L^{p-1} R$-cycle.
}
\label{fig:combo2}
\end{figure}

The dynamics are summarised by Fig.~\ref{fig:combo2}
which shows four representative $(\tau_L,\tau_R)$-two-parameter bifurcation diagrams
and shows how $Q_4$, for the pairs $(\delta_L,\delta_R)$, can be divided into different subsets
based on features of the bifurcation diagrams.
If $-1 < \delta_R < 0$
the bifurcation diagram has a (red) horizontal strip
$-\delta_R - 1 < \tau_R < \delta_R + 1$
where the right half-map has a stable fixed point.
Otherwise the dominate periodic solutions are those
whose symbolic itinerary contains exactly one $R$.
That is, for a period-$p$ solution the itinerary is $L^{p-1} R$
(or a cyclic permutation of this).
Its stability multipliers are the eigenvalues of $M_p = A_R A_L^{p-1}$.
Notice $\det(M_p) = \delta_L^{p-1} \delta_R < 0$ (because $\delta_R < 0 < \delta_L$).
So if $\det(M_p) > -1$ then the $L^{p-1} R$-cycle is stable when
\begin{equation}
-\det(M_p)-1 < {\rm trace}(M_p) < \det(M_p)+1.
\end{equation}
Regions where $L^{p-1} R$-cycles are stable and admissible are coloured
in Fig.~\ref{fig:combo2}.
Roughly speaking these regions are narrow and aligned vertically.
Their left boundary is where ${\rm trace}(M_p) = \det(M_p)+1$,
where $M_p$ has an eigenvalue $1$,
and their right boundary is where ${\rm trace}(M_p) = -\det(M_p)-1$,
where $M_p$ has an eigenvalue $-1$.
Upper boundaries are where admissibility is lost via
a border-collision bifurcation where one point of the $L^{p-1} R$-cycle
is located on the switching manifold.

As the pair $(\delta_L,\delta_R)$ is varied,
the left and right boundaries coincide,
and the stability region vanishes, when $\det(M_p) = -1$.
That is,
\begin{equation}
\delta_R = \frac{-1}{\delta_L^{p-1}}.
\end{equation}
These are curves labelled $\gamma_p$ in Fig.~\ref{fig:combo2}.
As $p \to \infty$ the curves converge to $\delta_L = 1$ and part of the $\delta_L$-axis.

Thus with $0 < \delta_L < 1$ and $-1 < \delta_R < 0$,
for each $p \ge 2$ the $(\tau_L,\tau_R)$-bifurcation diagram has a region where
there exists a stable $L^{p-1} R$-cycle.
The top-left bifurcation diagram of Fig.~\ref{fig:combo2} provides a typical example.
Only finitely many regions are visible because as $p$ increases
they only exist for successively larger (more negative) values of $\tau_R$.
By then increasing the value of $\delta_L$, but keeping $\delta_R$ fixed,
the regions vanish as we cross the curves $\gamma_p$ for successively decreasing values of $p$.
The top-right bifurcation diagram of Fig.~\ref{fig:combo2}
is for $(\delta_L,\delta_R)$ between $\gamma_3$ and $\gamma_4$,
so the period-$p$ regions exist for $p=2$ and $p=3$ but no higher values of $p$.

If instead we fix $\delta_R < -1$ and start with $\delta_L$ such that
$(\delta_L,\delta_R)$ lies to the left of $\gamma_2$, then again the period-$p$ region
exists for all $p \ge 2$ (bottom-left bifurcation diagram).
But now as we increase the values of $\delta_L$
the regions vanish for successively increasing values of $p$.
With $\delta_L > 1$ and $\delta_R < -1$, as in the bottom-right bifurcation diagram,
stable, admissible periodic solutions cannot exist and all attractors appear to be chaotic.
Here the region where an attractor exists
has a similar spiked shape to the region of \cite{Si20d}
where the origin is asymptotically stable when $\mu = 0$.
This is because asymptotic stability of the origin with $\mu = 0$
implies the existence of a local attractor for all $\mu \ne 0$ \cite{Si20b}.

In the bifurcation diagrams, just past the right boundary of a period-$p$ region
where the $L^{p-1} R$-cycle attains an eigenvalue of $-1$,
there can exist a region where a period-$2p$ solution is stable.
For example in the brown region of the top-right bifurcation diagram of Fig.~\ref{fig:combo2},
there exists a stable $LRRR$-cycle.

\begin{figure}[htbp]
\centering
\begin{subfigure}[b]{.5\linewidth}
    \centering
    \caption{}
    \includegraphics[width=.99\textwidth]{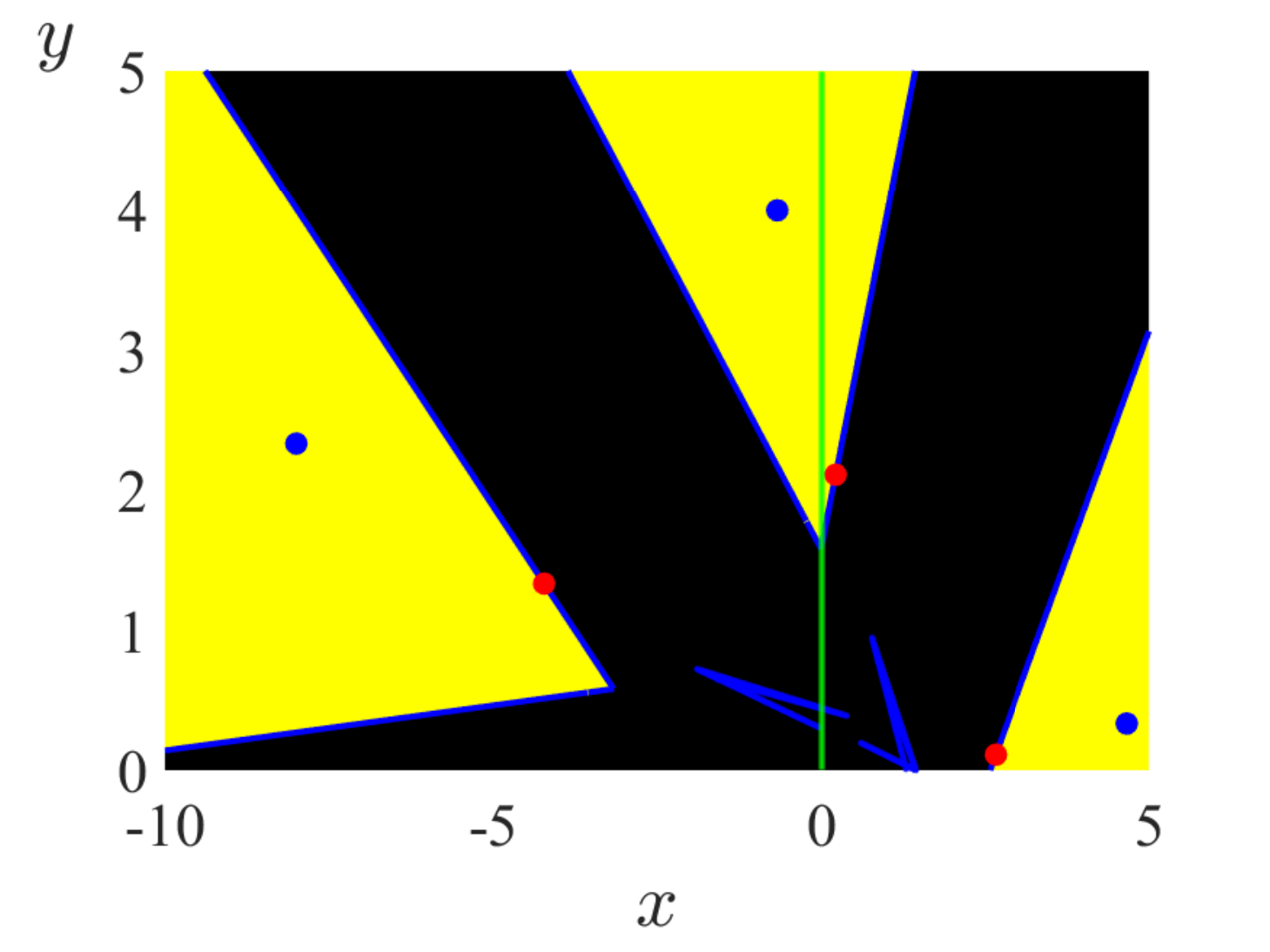}
  \end{subfigure}%
  \begin{subfigure}[b]{.5\linewidth}
    \centering
    \caption{}
    \includegraphics[width=.99\textwidth]{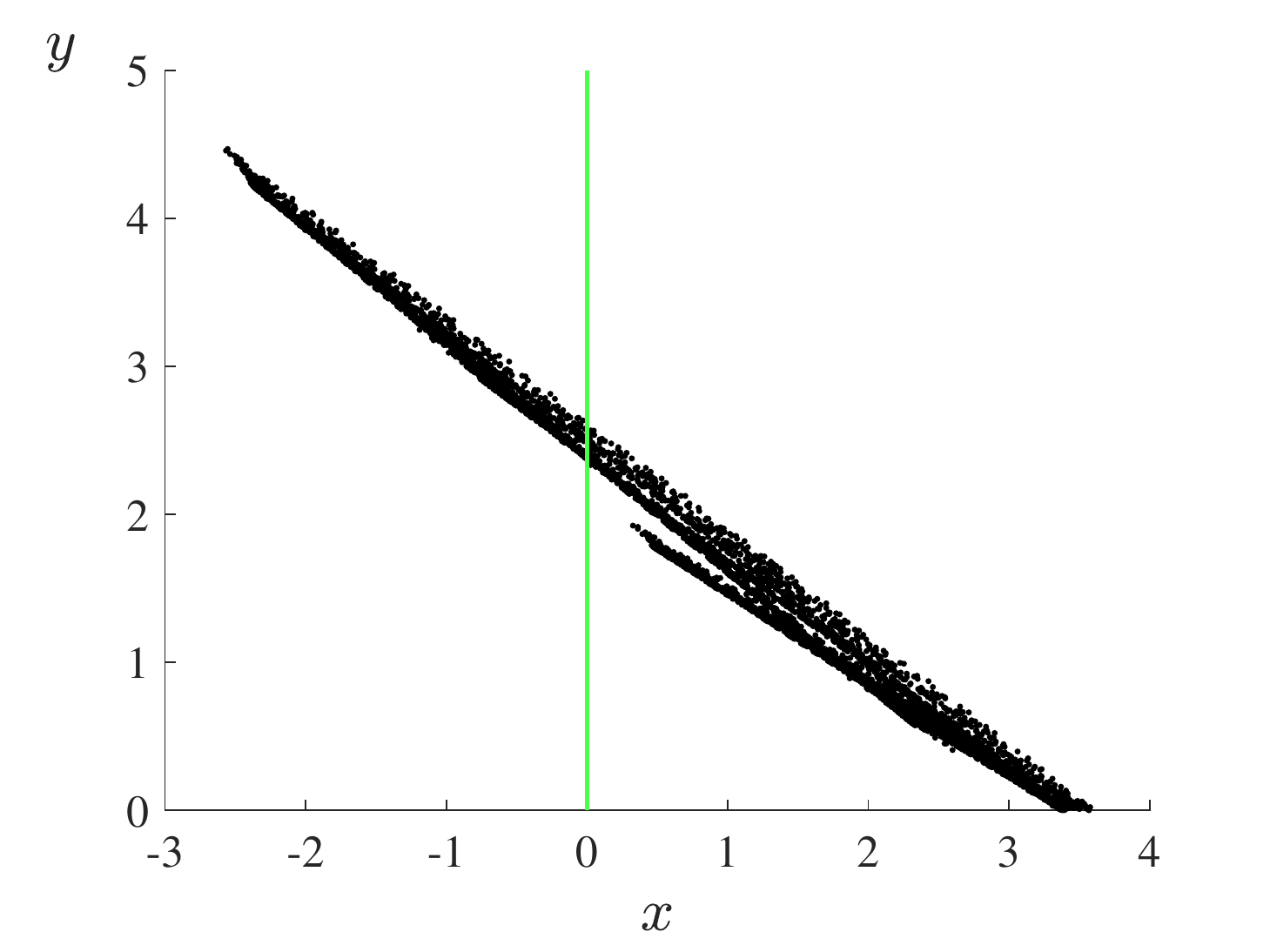}
  \end{subfigure}%  
  \caption{
  Panel (a) is a phase portrait of \eqref{eq:bcnf} with
  $(\tau_L,\delta_L,\tau_R,\delta_R) = (0.5,0.5,-2,-0.5)$.
  There exists a stable $LLR$-cycle (blue circles)
  and an apparently chaotic attractor (also blue).
  The stable manifold of a saddle $LRR$-cycle
  forms the boundary between their basins of attraction.
  Panel (b) is a phase portrait of \eqref{eq:bcnf} with
  $(\tau_L,\delta_L,\tau_R,\delta_R) = (2,0.75,-1,-1.25)$
  corresponding to Fig.~11 of \cite{MiRa96}.
    }
  \label{fig:positivePhasePortraits1}
\end{figure}

In the black regions of the bifurcation diagrams
the attractor is typically chaotic.
Attractors can coexist and Fig.~\ref{fig:positivePhasePortraits1}(a) shows a typical example.
Here the boundary between two basins of attraction
is given by the stable manifold of a saddle $LRR$-cycle.

Chaotic attractors undergo several global bifurcations as we move about the black regions.
These are described in \cite{MiRa96} for a one-parameter family.
Fig.~\ref{fig:positivePhasePortraits1}(b) shows an example of an attractor in their family
(in terms of the normal for \eqref{eq:bcnf}).
They found the attractor can be bounded by images of the switching manifold
and sometimes also by one or more unstable manifolds.
Bifurcations occur as the geometry of these one-dimensional objects changes.

%===============================================================================
\section{Application to power converters}
\label{sec:powerConverters}

Whenever an electrical device requires the use of current with a voltage different to that
of its power supply, it is necessary to alter the voltage level.
This is usually achieved with a power converter.
Power converters commonly employ rapid switching to minimize
energy loss and allow components (such as inductors and capacitors) to be small in size.

Here we consider the boost converter model of Deane \cite{De92c}.
This model has two variables: $u(t)$ and $v(t)$ for the current and voltage, respectively, in an inductor.
To regulate the output voltage the converter has a switch that closes
whenever $\frac{t}{T} \in \mathbb{Z}$, where $t \in \mathbb{R}$ is time
and $T > 0$ is the period of the converter.
While the switch is closed the variables evolve according to
\begin{align}
        \dot{u} &= \frac{V_I}{L}, &
        \dot{v} &= -\frac{1}{R C} \,v,
    \label{eq:closedODEs}
\end{align}
and Deane \cite{De92c} gives the following parameter values
\begin{align}
R &= 20, &
L &= 0.001, &
C &= 0.000012, &
T &= 0.0001, &
V_I &= 10.
\label{eq:powerConverterParameters}
\end{align}
The switch opens whenever the value of the current $u(t)$ reaches a threshold
value $I_{\rm ref}$, and while the switch is open the variables evolve according to
\begin{align}
        \dot{u} &= \frac{V_I}{L} - \frac{1}{L} \,v, &
        \dot{v} &= \frac{1}{C} \,u - \frac{1}{R C} \,v.
    \label{eq:openODEs}
\end{align}

\begin{figure}[htbp]
\centering
\includegraphics[scale=0.35]{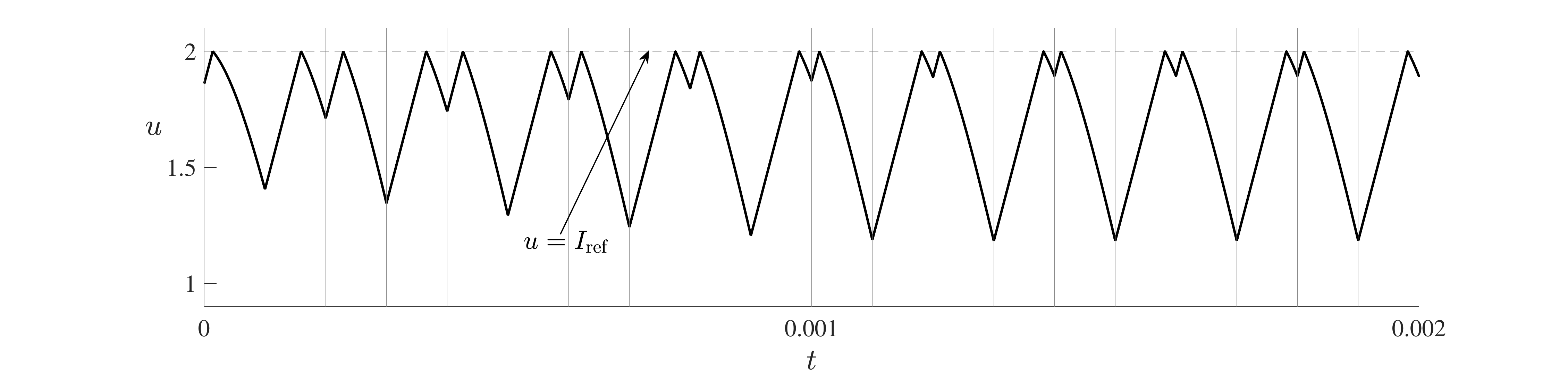}
\caption{
A time series of the boost converter model with \eqref{eq:powerConverterParameters} and $I_{\rm ref} = 2$.
A switch in the converter closes whenever $\frac{t}{T} \in \mathbb{Z}$ (vertical grey lines)
after which the current $u(t)$ increases.
The switch opens whenever $u(t) = I_{\rm ref}$ after which the current decreases.
}
\label{fig:timeseries}
\end{figure}

Fig.~\ref{fig:timeseries} shows a typical time series.
Notice the system evolves according to \eqref{eq:closedODEs} (switch closed)
until $u(t) = I_{\rm ref}$, then evolves according to \eqref{eq:openODEs}
until $t = n T$ for some $n \in \mathbb{Z}$.
For the given parameter values the dynamics of the converter converges
to a period-$2 T$ orbit.

\begin{figure}[htbp]
\centering
\includegraphics[scale=0.5]{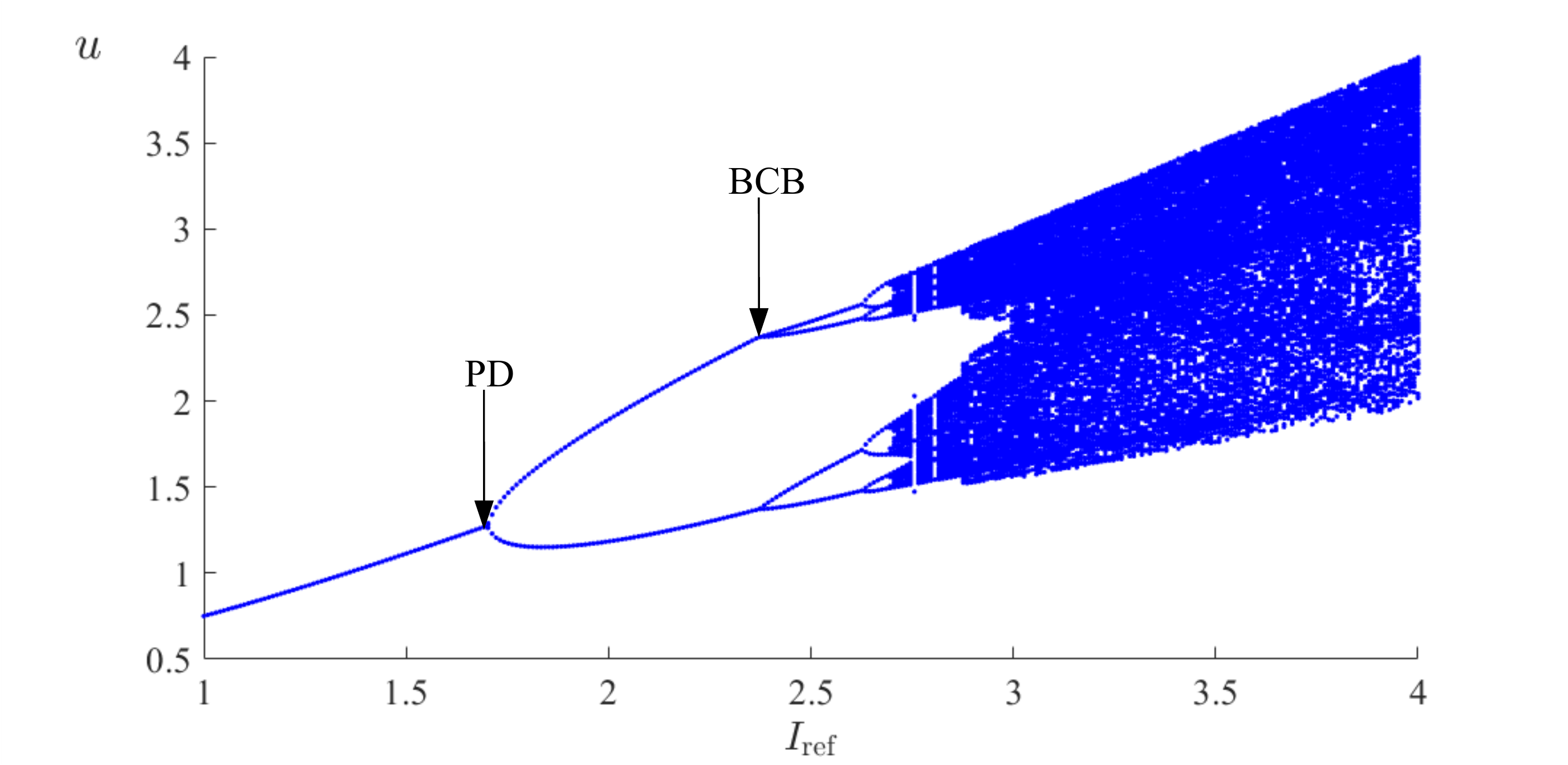}
\caption{
A numerically computed bifurcation diagram of the boost converter model
with \eqref{eq:powerConverterParameters}
(PD: period-doubling bifurcation; BCB: border-collision bifurcation).
}
\label{fig:boostBifDiag}
\end{figure}

Fig.~\ref{fig:boostBifDiag} shows a bifurcation diagram obtained by varying the value of $I_{\rm ref}$.
For small values of $I_{\rm ref}$ solutions converge to a period-$T$ orbit.
This solution loses stability in a period-doubling bifurcation at $I_{\rm ref} \approx 1.7060$
(labelled PD in Fig.~\ref{fig:boostBifDiag})
past which solutions converge to a period-$2T$ orbit.
The period doubles again at $I_{\rm ref} \approx 2.3721$,
except this is not another period-doubling bifurcation,
it is a border-collision bifurcation (labelled BCB) that mimics supercritical period-doubling.
In the remainder of this section we analyse this bifurcation
by using the results of the previous sections
and explain how it can differ for different values of the parameters \eqref{eq:powerConverterParameters}.

To analyse the border-collision bifurcation, as in \cite{De92c} we construct a stroboscopic map.
Given values $u$ and $v$ for the variables at a time $t = n T$, where $n \in \mathbb{Z}$,
let $(u',v') = g(u,v)$ be the values of variables at $t = (n+1) T$.
The map $g$ is piecewise-smooth, specifically it has the form
\begin{equation}
g(u,v) = \begin{cases}
g_L(u,v), & u \le I_{\rm ref} - \frac{V_I T}{L}, \\
g_R(u,v), & u \ge I_{\rm ref} - \frac{V_I T}{L},
\end{cases}
\label{eq:stroboscopicMap}
\end{equation}
because the switch may or not open between the times $n T$ and $(n+1) T$.
If $u < I_{\rm ref} - \frac{V_I T}{L}$
the switch is closed for the whole period
and from the flow of \eqref{eq:closedODEs} we obtain
\begin{equation}
g_L(u,v) = \begin{bmatrix}
u + \frac{V_I T}{L} \\
{\rm e}^{-\frac{T}{R C}} v
\end{bmatrix}.
\label{eq:gL}
\end{equation}
If $u > I_{\rm ref} - \frac{V_I T}{L}$
the switch opens at $t = n T + t_{\rm sw}$ where
$t_{\rm sw} = \frac{L}{V_I} (I_{\rm ref} - u)$.
At this time the current and voltage are given by \eqref{eq:gL}
except with $t_{\rm sw}$ in place of $T$.
By then evolving these values under the flow of \eqref{eq:openODEs}
(readily obtainable because \eqref{eq:openODEs} is affine) we obtain
(after much simplification)
\begin{equation}
g_R(u,v) =
{\rm e}^{-k t}
\begin{bmatrix}
\left( I_{\rm ref} - \frac{V_I}{R} \right) \cos(\omega t)
+ \frac{1}{\omega L} \left( V_I - {\rm e}^{-2 k t_{\rm sw}} v + k L \left( I_{\rm ref} - \frac{V_I}{R} \right)
\right) \sin(\omega t) \\
\frac{1}{\omega C} \left( I_{\rm ref} - \frac{V_I}{R} \right) \sin(\omega t)
+ \left( V_I - {\rm e}^{-2 k t_{\rm sw}} v \right)
\left( \frac{k}{\omega} \sin(\omega t) - \cos(\omega t) \right)
\end{bmatrix} +
\begin{bmatrix}
\frac{V_I}{R} \\ V_I
\end{bmatrix},
\label{eq:gR}
\end{equation}
where $k = \frac{1}{2 R C}$,
$\omega = \sqrt{\frac{1}{L C} - k^2}$,
and $t = T - t_{\rm sw}$.
These formulas are also given in \cite{De92c}.

The period-$2T$ orbit described above corresponds to a period-two solution ($RR$-cycle) of $g$.
The border-collision bifurcation occurs when one point of this solution
collides with the switching manifold of $g$.
Thus in a neighbourhood of this point the second iterate of $g$, call it $f$, has the form
\begin{equation}
f(u,v) = \begin{cases}
(g_R \circ g_L)(u,v), & u \le I_{\rm ref} - \frac{V_I T}{L}, \\
(g_R \circ g_R)(u,v), & u \ge I_{\rm ref} - \frac{V_I T}{L}.
\end{cases}
\label{eq:secondIterate}
\end{equation}
We can then bring \eqref{eq:secondIterate}
into the general form \eqref{eq:fGeneral} via a change of variables.
To end up with $\delta_R < 0 < \delta_L$ (matching Sections \ref{sec:muNegative} and \ref{sec:muPositive})
we assume this change of variables is done so that $g_R \circ g_R$ becomes the left piece of \eqref{eq:fGeneral}
and $g_R \circ g_L$ becomes the right piece of \eqref{eq:fGeneral}.

Since we have explicit formulas for $g_L$ and $g_R$,
it is a straight-forward (although tedious)
exercise to evaluate the Jacobian matrices ${\rm D}(g_R \circ g_R)$
and ${\rm D}(g_R \circ g_L)$ at the border-collision bifurcation.
Numerically we find that with the parameter values \eqref{eq:powerConverterParameters},
at the border-collision bifurcation the trace and determinant of ${\rm D}(g_R \circ g_R)$ are
\begin{align}
\tau_L &= -0.2726, &
\delta_L &= 0.1136, 
\end{align}
and the trace and determinant of ${\rm D}(g_R \circ g_L)$ are
\begin{align}
\tau_L &= -1.4992, &
\delta_L &= -0.2222, 
\end{align}
to four decimal places.
With these values the border-collision normal form
has a stable fixed point for $\mu < 0$ and a stable $LR$-cycle for $\mu > 0$.
Specifically for $\mu < 0$, $(\delta_L,\delta_R)$ belongs to subset (vii) of Fig.~\ref{fig:combo1}
and $(\tau_L,\tau_R)$ belongs to the red strip $-\delta_L - 1 < \tau_L < \delta_L + 1$
where the left half-map has a stable fixed point.
For $\mu > 0$, $(\delta_L,\delta_R)$ belongs to the top-left part of $Q_4$ and
$(\tau_L,\tau_R)$ belongs to the cyan region where there exists a stable $LR$-cycle.
In terms of the stroboscopic map $g$,
this corresponds to a stable period-two solution bifurcating to a stable period-four solution
as the value of $I_{\rm ref}$ is increased to pass through the border-collision bifurcation,
which confirms what we are seeing in Fig.~\ref{fig:boostBifDiag}.

\begin{figure}[htbp]
\centering
\includegraphics[width=.5\textwidth]{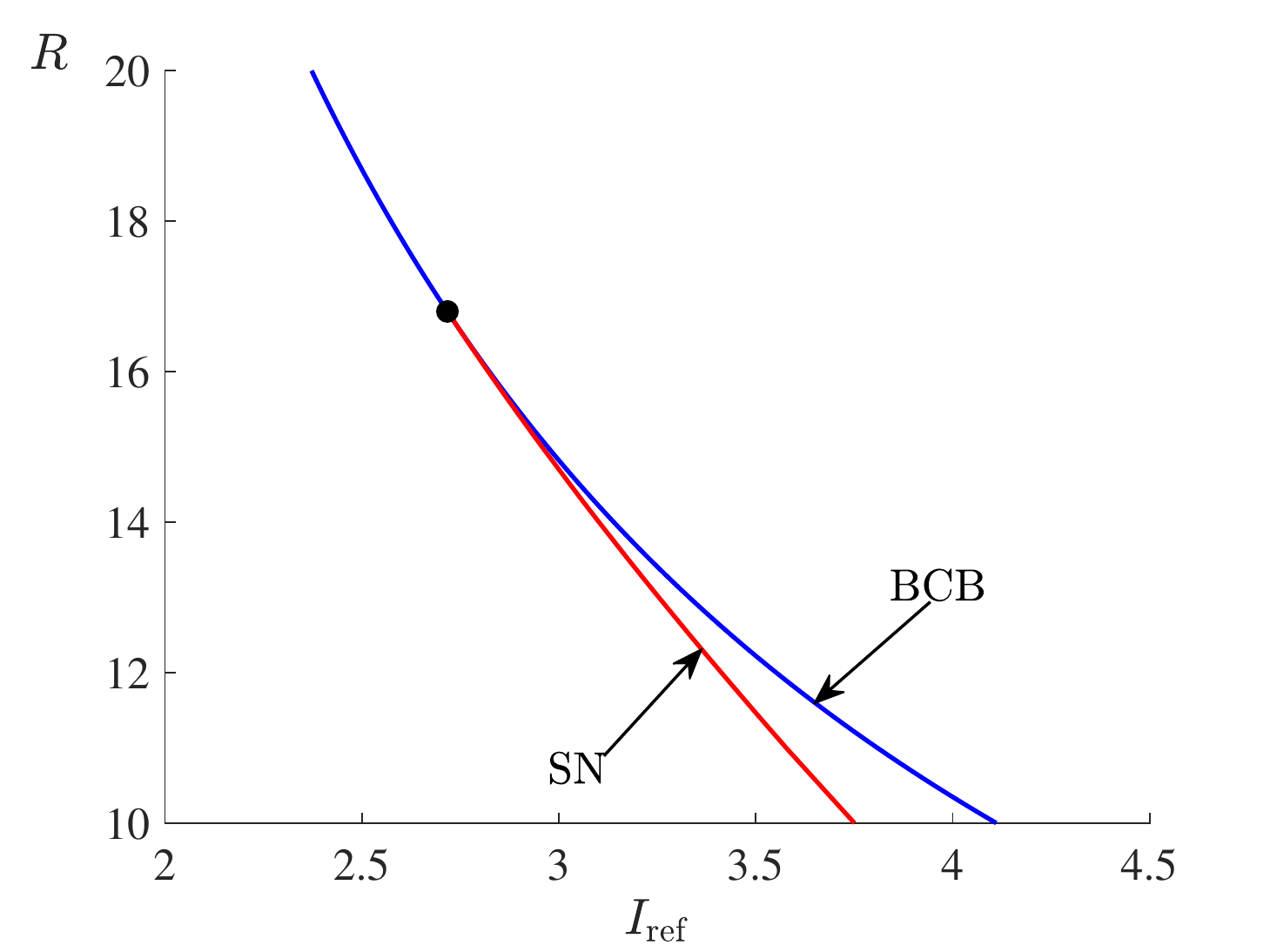}
\caption{
A two-parameter bifurcation diagram of the boost converter model
obtained by varying $I_{\rm ref}$ and $R$
and keeping all other parameter values fixed as in \eqref{eq:powerConverterParameters}
(SN: saddle-node bifurcation; BCB: border-collision bifurcation).
}
\label{fig:boostBifSet}
\end{figure}

\begin{figure}[htbp]
\centering
\includegraphics[width=.6\textwidth]{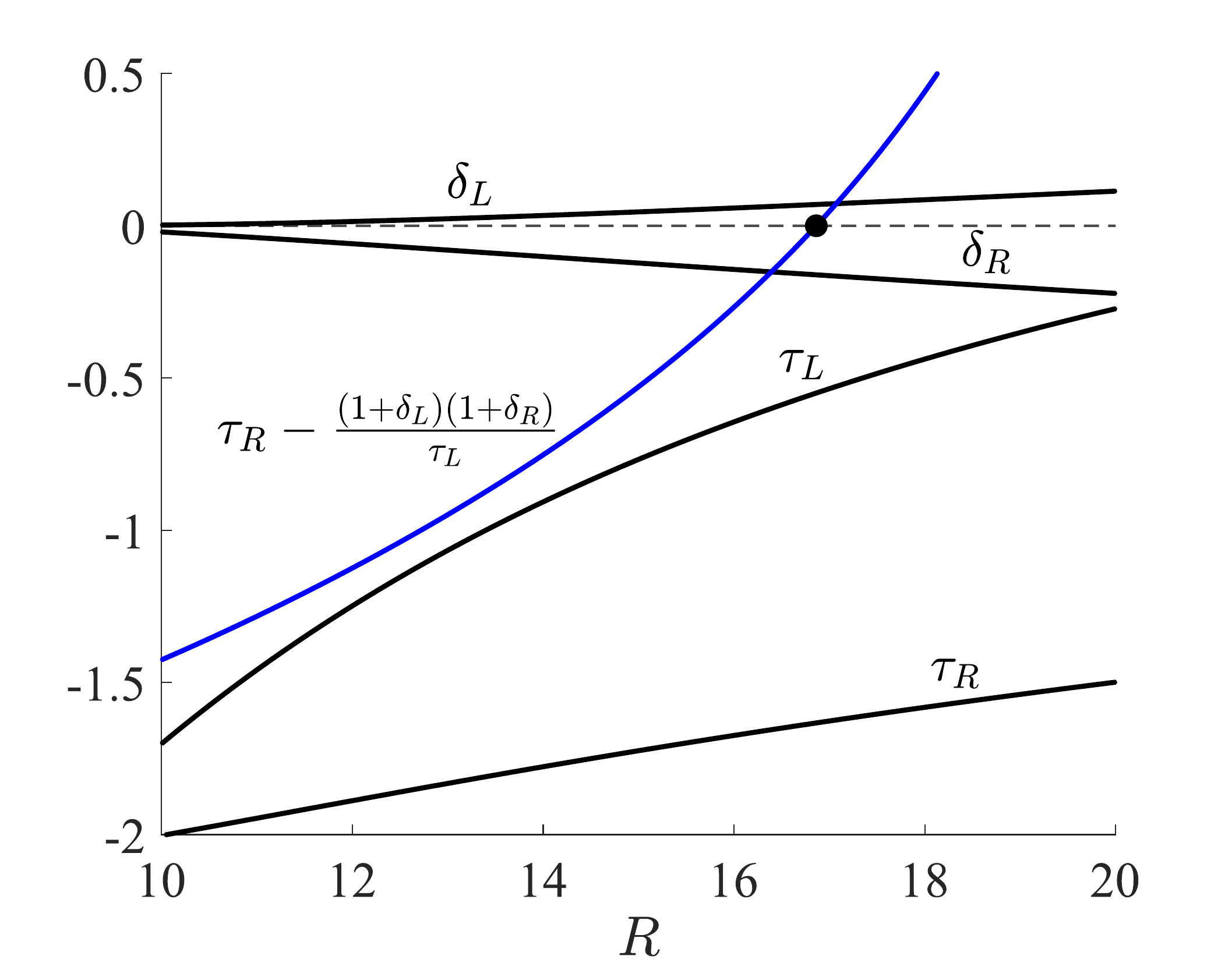}
\caption{
The values of $\tau_L$, $\delta_L$, $\tau_R$, and $\delta_R$
at the border-collision bifurcation of the boost converter.
The parameter values are given by \eqref{eq:powerConverterParameters} except $R$ is variable.
The value of $I_{\rm ref}$ is that at which the border-collision bifurcation occurs, see Fig.~\ref{fig:boostBifSet}.
We have also plotted the quantity $\tau_R - \frac{(1+\delta_L)(1+\delta_R)}{\tau_L}$.
The sign of this quantity governs the criticality of the border-collision bifurcation
(in the context of the period-$4T$ orbit).
}
\label{fig:R-params}
\end{figure}

We now consider parameter values different to \eqref{eq:powerConverterParameters}.
As we decrease the value of the resistance $R$ from the value $20$
and keep the other values in \eqref{eq:powerConverterParameters} fixed,
a border-collision bifurcation still occurs but at a varying value of $I_{\rm ref}$, as shown in Fig.~\ref{fig:boostBifSet}.
The values of $\tau_L$, $\delta_L$, $\tau_R$, and $\delta_R$ also
vary with $R$, as shown in Fig.~\ref{fig:R-params}.
These values belong to the same subsets and regions of Figs.~\ref{fig:combo1} and \ref{fig:combo2}
until at $R \approx 16.8623$ we exit the cyan region for $\mu > 0$.
Specifically we cross the left boundary $\tau_R = \frac{(1+\delta_L)(1+\delta_R)}{\tau_L}$
where the $LR$-cycle attains an eigenvalue of $1$.
Beyond this boundary the border-collision normal form has no local attractor for $\mu > 0$,
and an unstable $LR$-cycle is created for $\mu < 0$.

In terms of $g$, the border-collision bifurcation now mimics subcritical period-doubling
because an unstable period-four solution is created and grows as the value of $I_{\rm ref}$ decreases.
However this solution soon becomes stable in a saddle-node bifurcation.
In the two-parameter bifurcation diagram, Fig.~\ref{fig:boostBifSet},
the curve of saddle-node bifurcations emanates from the curve of border-collision bifurcations
with a quadratic tangency at the codimension-two point ($R \approx 16.8623$)
in accordance with the bifurcation theory for such points \cite{CoDe10,Si10}.

Thus the power converter has a small region of bistability
(between the saddle-node and border-collision bifurcations)
where stable period-$2 T$ and period-$4 T$ orbits coexist.
Below the codimension-two point and to the right of the border-collision bifurcation,
the power converter model has a stable period-$4 T$ orbit
whereas the corresponding border-collision normal form has no attractor
because here the period-$4 T$ orbit is a consequence of nonlinearities in $g$
that are not captured by the normal form.

We explored the effect of varying the other model parameters from their values in \eqref{eq:powerConverterParameters}
and found that the corresponding values of $\tau_L$, $\delta_L$, $\tau_R$, and $\delta_R$
either remained within the subsets and regions of Figs.~\ref{fig:combo1} and \ref{fig:combo2} described above,
or exited the cyan region by crossing its left boundary.
This suggests that over a wide range of parameter values,
the border-collision bifurcation of the power converter only mimics period-doubling,
but the doubling can be supercritical or subcritical depending on the parameters.

%===============================================================================
\section{Discussion}
\label{sec:conc}

Border-collision bifurcations occur in mathematical models of diverse physical phenomena.
Subject to reasonable genericity conditions, in two dimensions the leading-order dynamics are characterised by four scalar parameters, $\tau_L$, $\tau_R$, $\delta_L$, and $\delta_R$, see \S\ref{sec:bcbs}.
In this paper we have attempted to provide a high-level overview of the dynamics of normal form in the non-invertible case, $\delta_L \delta_R < 0$.
To grasp four-dimensional parameter space, we have considered two-dimensional slices (two-parameter bifurcation diagrams) defined by fixing the values of $\delta_L > 0$ and $\delta_R < 0$,
and studied how the dynamics across the slices is different with different values.
With instead $\delta_L < 0$ and $\delta_R > 0$
the same dynamical transitions occur
when the sign of the border-collision bifurcation parameter $\mu$ is reversed.

The results are summarised by Figs.~\ref{fig:combo1} and \ref{fig:combo2}.
The two figures correspond to different sides of the border-collision bifurcation.
The intention is for these to be applied to border-collision bifurcations in models
to make predictions about what dynamics are created
and how the dynamics differs for different parameter values.

We illustrated this in \S\ref{sec:powerConverters} for a classic boost converter model.
The parameter values of \cite{De92c}
correspond a red region (stable fixed point) in Fig.~\ref{fig:combo1} 
and a cyan region (stable $LR$-cycle) in Fig.~\ref{fig:combo2}.
Thus the bifurcation in the converter affects a doubling of the period.
As we vary the parameter values of the converter, we move about Figs.~\ref{fig:combo1} and \ref{fig:combo2}.
Numerically we found we remained in the red region
but could exit the cyan region by passing through a boundary where the $LR$-cycle loses stability.
This has the effect of flipping the criticality of the period-doubling-like bifurcation in the converter.

%===============================================================================
\appendix
\section{Existence of period-3 regions} % for $\mu < 0$}
\label{sec:A1}

\begin{figure}[htbp]
\centering
     \includegraphics[width=.4\textwidth]{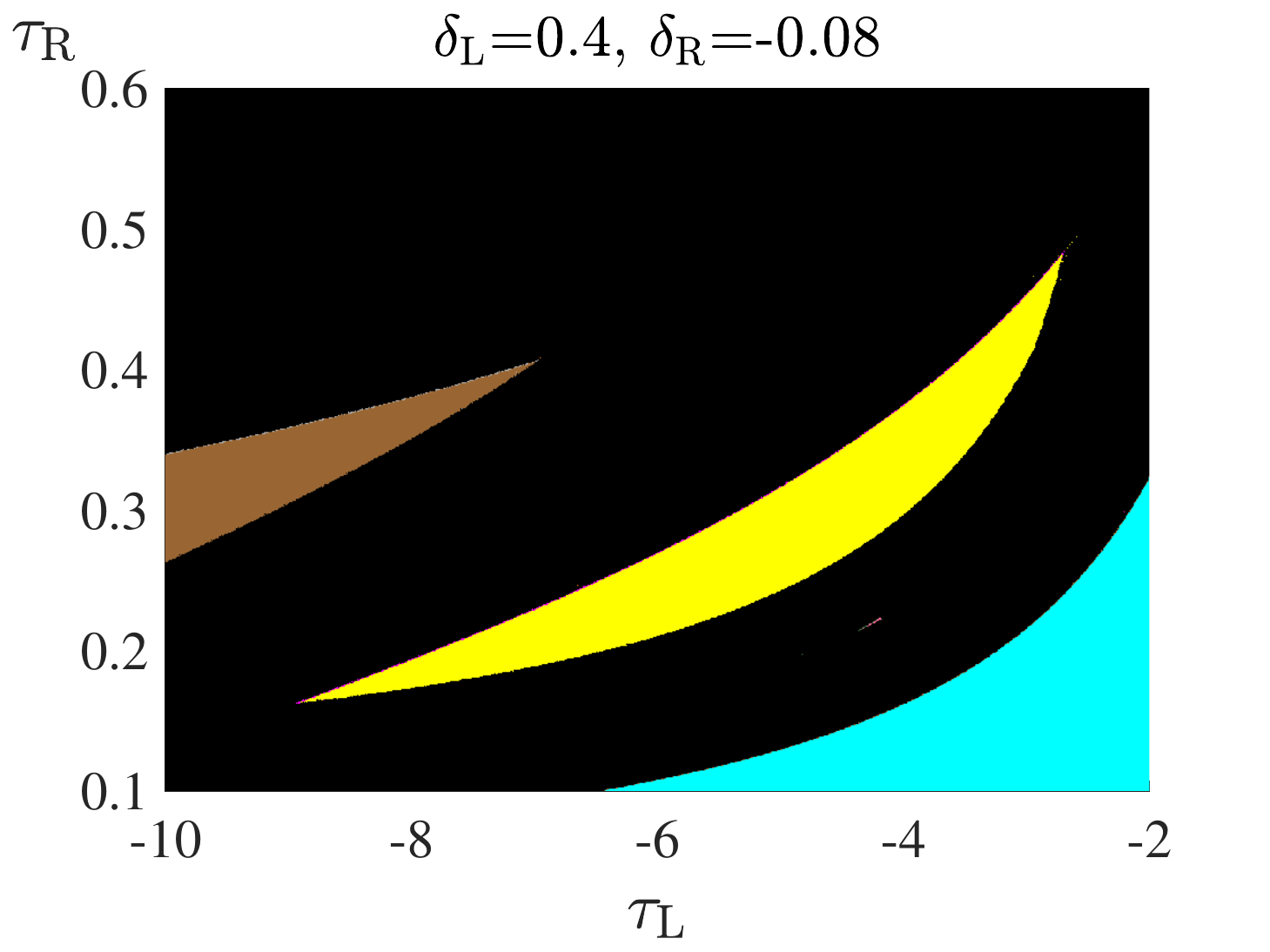}
  \caption{
  A two-parameter bifurcation diagram repeating Fig.~\ref{fig:combo1}(i)
  but over a smaller range of $\tau_L$ and $\tau_R$ values.
  In the yellow region the map with $\mu < 0$ has a stable $LRR$-cycle.
  This region has two boundaries given by \eqref{eq:tauL1} and \eqref{eq:tauL2}.
  }
  \label{fig:zooom}
  \end{figure}
  
Here we derive the curve $\zeta_3$ that forms
the boundary between subsets (i) and (vii) of Fig.~\ref{fig:combo1}.
For any $(\delta_L,\delta_R)$ in subset (i),
the $(\tau_L,\tau_R)$-bifurcation diagram has two regions
where there exists a stable $LRR$-cycle, Fig.~\ref{fig:combo1}(i).
The curve $\zeta_3$ is where the upper region, shown more clearly in Fig.~\ref{fig:zooom} shrinks to a point.

The upper region has two boundaries.
The left boundary is where the $LRR$-cycle loses stability by attaining an eigenvalue of $-1$.
Its eigenvalues are those of $M = A_L A_R^2$,
so an eigenvalue of $-1$ occurs when ${\rm det}(M)+{\rm trace}(M)+1=0$.
By directly evaluating and rearranging this equation we find that the left boundary is given by
%\begin{align}
%\label{eq:eig_mat}
%M=A_{L}A_{R}^{2}=\begin{pmatrix}
%\tau_{T}^2\tau_{L}-\tau_{R}\delta_{L}-\tau_{R}\delta_{R} & \tau_{R}\tau_{L}-\delta_{L}\\
%-\tau_{R}\tau_{L}\delta_{R} & -\tau_{L}\delta_{R}
%\end{pmatrix}.
%\end{align}
\begin{equation}
\label{eq:tauL1}
\tau_{L}=\frac{\tau_{R}(\delta_{R}+\delta_{L})-\delta_{R}^{2}\delta_{L}-1}{\tau_{R}^{2}-\delta_{R}}.
\end{equation}
The right boundary is where the
$LRR$-cycle undergoes a border-collision bifurcation
and becomes virtual by having one of its points collide with the switching manifold.
By direct calculations, or by using the methods in \cite{Si16},
it is a simple exercise to show that the bifurcation occurs when ${\rm det}(I + A_{R}+A_{R}A_{L})=0$.
By evaluating this equation we obtain the formula
%\begin{equation}
%  I + A_{R}+A_{R}A_{L}=\begin{pmatrix}
%1+\tau_{R}+\tau_{R}\tau_{L}-\delta_{L} & 1+\tau_{R}\\
%-\delta_{R}-\tau_{L}\delta_{R} & 1-\delta_{R}
%\end{pmatrix}.
%\end{equation}
\begin{equation}
\label{eq:tauL2}
\tau_{L}=\frac{\delta_{L}(1-\delta_{R})-\tau_{R}-1}{\tau_{R}+\delta_{R}},
\end{equation}
for the right boundary of the upper region.

The two vertices of this region
(the tips of the `horns' of the crescent) satisfy both \eqref{eq:tauL1} and \eqref{eq:tauL2}.
By equating these we can eliminate $\tau_L$ to obtain
\begin{equation}
a \tau_R^3 + b \tau_R^2 + c \tau_R + d = 0,
\label{eq:cubic}
\end{equation}
where
\begin{align*}
a &= 1, \\
b &= \delta_L \delta_R + \delta_R + 1, \\
c &= -\delta_L \delta_R^2 + \delta_L \delta_R + \delta_R^2 - \delta_R - 1, \\
d &= -\delta_L \delta_R^3 - \delta_L \delta_R^2 + \delta_L \delta_R - 2 \delta_R \,.
\end{align*}
That is, the $\tau_R$-values of the vertices satisfy the cubic \eqref{eq:cubic}.
The region vanishes when the two vertices coincide,
thus when the discriminant of \eqref{eq:cubic} is zero.
The discriminant of \eqref{eq:cubic} is
\begin{equation}
\Delta = 18 a b c d - 4 b^3 d + b^2 c^2 - 4 a c^3 - 27 a^2 d^2,
\end{equation}
therefore the curve $\zeta_3$ is given implicitly by
the two-variable polynomial equation $\Delta = 0$.
This curve is shown in Fig.~\ref{fig:combo1}.
As $\delta_L$ is varied from $0$ to $1$,
it appears to decrease monotonically from
$\delta_R \approx -0.08238$ to $\delta_R = -1$.

%=====================================================================
\section*{Acknowledgements}

The authors were supported by Marsden Fund contract MAU1809,
managed by Royal Society Te Ap\={a}rangi.


\begin{thebibliography}{10}

\bibitem{Me07}
J.D. Meiss.
\newblock {\em Differential Dynamical Systems.}
\newblock SIAM, Philadelphia, 2007.

\bibitem{DiBu08}
M.~di~Bernardo, C.J. Budd, A.R. Champneys, and P.~Kowalczyk.
\newblock {\em Piecewise-smooth Dynamical Systems. Theory and Applications.}
\newblock Springer-Verlag, New York, 2008.

\bibitem{Di03}
M.~di~Bernardo.
\newblock Normal forms of border collision in high dimensional non-smooth maps.
\newblock In {\em Proceedings IEEE ISCAS, Bangkok, Thailand}, volume~3, pages
  76--79, 2003.

\bibitem{NuYo92}
H.E. Nusse and J.A. Yorke.
\newblock Border-collision bifurcations including ``period two to period
  three'' for piecewise smooth systems.
\newblock {\em Phys. D}, 57:39--57, 1992.

\bibitem{Si16}
D.J.W. Simpson.
\newblock Border-collision bifurcations in $\mathbb{R}^n$.
\newblock {\em SIAM Rev.}, 58(2):177--226, 2016.

\bibitem{Lo78}
R.~Lozi.
\newblock Un attracteur \'{e}trange(?) du type attracteur de {H}\'{e}non.
\newblock {\em J. Phys. (Paris)}, 39(C5):9--10, 1978.
\newblock In French.

\bibitem{BaYo98}
S.~Banerjee, J.A. Yorke, and C.~Grebogi.
\newblock Robust chaos.
\newblock {\em Phys. Rev. Lett.}, 80(14):3049--3052, 1998.

\bibitem{GlSi21}
P.A. Glendinning and D.J.W. Simpson.
\newblock A constructive approach to robust chaos using invariant manifolds and
  expanding cones.
\newblock {\em Discrete Contin. Dyn. Syst.}, 41(7):3367--3387, 2021.

\bibitem{Si14}
D.J.W. Simpson.
\newblock Sequences of periodic solutions and infinitely many coexisting
  attractors in the border-collision normal form.
\newblock {\em Int. J. Bifurcation Chaos}, 24(6):1430018, 2014.

\bibitem{PuRo18}
A.~Pumari\~{n}o, J.A. Rodr{\'\i}guez, and E.~Vigil.
\newblock Renormalization of two-dimensional piecewise linear maps: {A}bundance
  of {2-D} strange attractors.
\newblock {\em Discrete Contin. Dyn. Syst.}, 38(2):941--966, 2018.

\bibitem{Si22c}
D.J.W.~Simpson.
\newblock Border-collision bifurcations from stable fixed points to
		any number of coexisting chaotic attractors.
\newblock {\em Submitted}, 2022. \texttt{arXiv:2207.10251}

\bibitem{BaGr99}
S.~Banerjee and C.~Grebogi.
\newblock Border collision bifurcations in two-dimensional piecewise smooth
  maps.
\newblock {\em Phys. Rev. E}, 59(4):4052--4061, 1999.

\bibitem{SiMe08b}
D.J.W. Simpson and J.D. Meiss.
\newblock Neimark-{S}acker bifurcations in planar, piecewise-smooth, continuous
  maps.
\newblock {\em SIAM J. Appl. Dyn. Sys.}, 7(3):795--824, 2008.

\bibitem{Si20e}
D.J.W. Simpson.
\newblock Detecting invariant expanding cones for generating word sets to
  identify chaos in piecewise-linear maps.
\newblock Submitted to: {\em J. Difference Eq. Appl.}
  \texttt{arXiv:2010.08241}, 2020.

\bibitem{ZhMo06b}
Z.T. Zhusubaliyev, E.~Mosekilde, S.~Maity, S.~Mohanan, and S.~Banerjee.
\newblock Border collision route to quasiperiodicity: Numerical investigation
  and experimental confirmation.
\newblock {\em Chaos}, 16(2):023122, 2006.

\bibitem{SuGa08}
I.~Sushko and L.~Gardini.
\newblock Center bifurcation for two-dimensional border-collision normal form.
\newblock {\em Int. J. Bifurcation Chaos}, 18(4):1029--1050, 2008.

\bibitem{MiRa96}
C.~Mira, C.~Rauzy, Y.~Maistrenko, and I.~Sushko.
\newblock Some properties of a two-dimensional piecewise-linear noninvertible
  map.
\newblock {\em Int. J. Bifurcation Chaos}, 6(12a):2299--2320, 1996.

\bibitem{Gl16e}
P.~Glendinning.
\newblock Bifurcation from stable fixed point to {2D} attractor in the border
  collision normal form.
\newblock {\em IMA J. Appl. Math.}, 81(4):699--710, 2016.

\bibitem{Ko05}
P.~Kowalczyk.
\newblock Robust chaos and border-collision bifurcations in non-invertible
  piecewise-linear maps.
\newblock {\em Nonlinearity}, 18:485--504, 2005.

\bibitem{SzOs09}
R.~Szalai and H.M. Osinga.
\newblock Arnol'd tongues arising from a grazing-sliding bifurcation.
\newblock {\em SIAM J. Appl. Dyn. Sys.}, 8(4):1434--1461, 2009.

\bibitem{De92c}
J.H.B. Deane.
\newblock Chaos in a current-mode controlled boost dc-dc converter.
\newblock {\em IEEE Trans. Circuits Systems I Fund. Theory Appl.},
  39(8):680--683, 1992.

\bibitem{DiBu01}
M.~di~Bernardo, C.J. Budd, and A.R. Champneys.
\newblock Normal form maps for grazing bifurcations in $n$-dimensional
  piecewise-smooth dynamical systems.
\newblock {\em Phys. D}, 160:222--254, 2001.

\bibitem{GuHo86}
J.~Guckenheimer and P.J. Holmes.
\newblock {\em Nonlinear Oscillations, Dynamical Systems, and Bifurcations of
  Vector Fields.}
\newblock Springer-Verlag, New York, 1986.

\bibitem{Ku04}
Yu.A. Kuznetsov.
\newblock {\em Elements of Bifurcation Theory.}, volume 112 of {\em Appl. Math.
  Sci.}
\newblock Springer-Verlag, New York, 3rd edition, 2004.

\bibitem{SiGl21}
D.J.W. Simpson and P.A. Glendinning.
\newblock Inclusion of higher-order terms in the border-collision normal form:
  persistence of chaos and applications to power converters.
\newblock Submitted to: {\em Phys. D}. \texttt{arXiv:2111.12222}, 2021.

\bibitem{Si20b}
D.J.W. Simpson.
\newblock Chaotic attractors from border-collision bifurcations: {S}table
  border fixed points and determinant-based {L}yapunov exponent bounds.
\newblock {\em NZJM}, 50:71--91, 2020.

\bibitem{DoKi08}
Y.~Do, S.D. Kim, and P.S. Kim.
\newblock Stability of fixed points placed on the border in the piecewise
  linear systems.
\newblock {\em Chaos Solitons Fractals}, 38(2):391--399, 2008.

\bibitem{Ga92}
L.~Gardini.
\newblock Some global bifurcations of two-dimensional endomorphisms by use of
  critical lines.
\newblock {\em Nonlinear Anal.}, 18(4):361--399, 1992.

\bibitem{Si20d}
D.J.W. Simpson.
\newblock The stability of fixed points on switching manifolds of
  piecewise-smooth continuous maps.
\newblock {\em J. Dyn. Diff. Equat.}, 32(3):1527--1552, 2020.

\bibitem{SiMe10}
D.J.W. Simpson and J.D. Meiss.
\newblock Resonance near border-collision bifurcations in piecewise-smooth,
  continuous maps.
\newblock {\em Nonlinearity}, 23(12):3091--3118, 2010.

\bibitem{Gl17b}
P.~Glendinning.
\newblock Less is more {I}: {A} pessimistic view of piecewise smooth
  bifurcation theory.
\newblock In A.~Colombo, M.~Jeffrey, J.~L\'{a}zaro, and J.~Olm, editors, {\em
  Extended Abstracts Spring 2016}, volume~8 of {\em Trends in Mathematics},
  pages 71--75. Birkh\"{a}user, Cham, 2017.

\bibitem{Ar88}
V.I. Arnol'd.
\newblock {\em Geometrical Methods in the Theory of Ordinary Differential
  Equations.}
\newblock Springer-Verlag, New York, 2nd edition, 1988.

\bibitem{ZhYa10}
Z.T. Zhusubaliyev, O.O. Yanochkina, E.~Mosekilde, and S.~Banerjee.
\newblock Two-mode dynamics in pulse-modulated control systems.
\newblock {\em Annual Rev. Control}, 34:62--70, 2010.

\bibitem{GaGa03}
M.~Gallegati, L.~Gardini, T.~Puu, and I.~Sushko.
\newblock Hicks' trade cycle revisited: Cycles and bifurcations.
\newblock {\em Math. Comput. Simulation}, 63:505--527, 2003.

\bibitem{LaMo06}
J.~Laugesen and E.~Mosekilde.
\newblock Border-collision bifurcations in a dynamic management game.
\newblock {\em Comput. Oper. Res.}, 33:464--478, 2006.

\bibitem{Ti02}
P.H.E. Tiesinga.
\newblock Precision and reliability of periodically and quasiperiodically
  driven integrate-and-fire neurons.
\newblock {\em Phys. Rev. E}, 65(4):041913, 2002.

\bibitem{SiMe09}
D.J.W. Simpson and J.D. Meiss.
\newblock Shrinking point bifurcations of resonance tongues for
  piecewise-smooth, continuous maps.
\newblock {\em Nonlinearity}, 22(5):1123--1144, 2009.

\bibitem{Fr90}
J.~Franks.
\newblock Periodic points and rotation numbers for area preserving
  diffeomorphisms of the plane.
\newblock {\em Inst. Hautes \'{E}tudes Sci. Publ. Math.}, 71:105--120, 1990.

\bibitem{Le01}
P.~Le~Calvez.
\newblock Rotation numbers in the infinite annulus.
\newblock {\em Proc. Amer. Math. Soc.}, 129(11):3221--3230, 2001.

\bibitem{HaZh18}
B.~Hao and W.~Zheng.
\newblock {\em Applied Symbolic Dynamics and Chaos.}
\newblock World Scientific, Singapore, 2nd edition, 2018.

\bibitem{HaWr08}
G.H. Hardy and E.M. Wright.
\newblock {\em An Introduction to the Theory of Numbers.}
\newblock Oxford University Press, New York, 6th edition, 2008.

\bibitem{CoDe10}
A.~Colombo and F.~Dercole.
\newblock Discontinuity induced bifurcations of non-hyperbolic cycles in
  nonsmooth systems.
\newblock {\em SIAM J. Appl. Dyn. Sys.}, 9(1):62--83, 2010.

\bibitem{Si10}
D.J.W. Simpson.
\newblock {\em Bifurcations in Piecewise-Smooth Continuous Systems.}, volume~70
  of {\em Nonlinear Science}.
\newblock World Scientific, Singapore, 2010.

\end{thebibliography}
\end{document}